\documentclass[aps, 
 amsmath,amssymb, amsbsy,
prb,
floatfix,
superscriptaddress,
twocolumn
]{revtex4-2}

\usepackage{gensymb}
\usepackage[caption=false]{subfig}
\usepackage{xcolor}
\usepackage{graphicx}
\usepackage{dcolumn}
\usepackage{bm}
\usepackage{float}
\usepackage{hyperref}
\usepackage{enumerate}
\usepackage{dsfont}
\usepackage{textcomp}
\usepackage{verbatim}

\usepackage{calligra} 
\DeclareMathAlphabet{\mathcalligra}{T1}{calligra}{m}{n} \DeclareFontShape{T1}{calligra}{m}{n}{<->s*[2.2]callig15}{} 

\newcommand*{\reverses}{\reflectbox{s}}%
\newcommand{\ch}[1]{\textcolor{black}{#1}}

\begin{document}

\title{Emergent Nonlocal Combinatorial Design Rules for Multimodal Metamaterials}

\author{Ryan van Mastrigt}
\email{r.vanmastrigt@uva.nl}
\affiliation{Institute of Physics, Universiteit van Amsterdam, Science Park 904, 1098 XH Amsterdam, The Netherlands}
\affiliation{AMOLF, Science Park 104, 1098 XG Amsterdam, The Netherlands}

\author{Corentin Coulais}%
\affiliation{Institute of Physics, Universiteit van Amsterdam, Science Park 904, 1098 XH Amsterdam, The Netherlands}

\author{Martin van Hecke}
\affiliation{AMOLF, Science Park 104, 1098 XG Amsterdam, The Netherlands}
\affiliation{Huygens-Kamerling Onnes Lab, Universiteit Leiden, Postbus 9504, 2300 RA Leiden, The Netherlands}


\date{\today}

\begin{abstract}{
Combinatorial mechanical metamaterials feature spatially textured soft modes 
that yield exotic and useful mechanical properties. While a single soft mode often can be 
rationally designed by following a set of tiling rules for the building blocks of the metamaterial,
it is an open question what design rules are required to realize multiple soft modes. 
Multimodal metamaterials would allow for advanced mechanical functionalities that can be 
selected on-the-fly. 
Here we introduce a transfer matrix-like framework to design multiple soft modes in combinatorial metamaterials
composed of aperiodic tilings of building blocks. 
We use this framework to derive rules for multimodal designs for
a specific family of building blocks. 
We show that such designs require a large number of degeneracies between constraints, and find precise rules
on the real space configuration that allow such degeneracies. 
These rules are significantly more complex than the simple tiling rules that emerge for single-mode metamaterials. For the specific example studied here, they can be expressed as local rules for tiles composed of pairs of building blocks in combination with 
a nonlocal rule in the form of a global constraint on the type of tiles that are allowed to appear together anywhere in the configuration. This nonlocal rule is
exclusive to multimodal metamaterials and exemplifies the complexity of rational design of multimode metamaterials. Our framework is a first step towards a systematic design strategy of multimodal metamaterials with spatially textured soft modes.
}
\end{abstract}

\maketitle
\section{Introduction}
The structure and proliferation of soft modes is paramount for understanding the mechanical properties of a wide variety of soft and flexible materials~\cite{alexander1998amorphous, hecke2009jamming, liu2010jamming, nicolas2018deformation, bertoldi2017flexible}.
Recently, computational and rational design of soft modes in designer matter has given rise to the field of mechanical metamaterials~\cite{sigmund1994materials, sigmund1997design, evans2000auxetic, reis2015designer, zadpoor2016mechanical, ion2016metamaterial, bertoldi2017flexible, zhu2020design, dykstra2022extreme}. Typically, such materials are structured such that a single soft mode controls the low energy deformations. Their geometric design is often based on that of a single zero-energy mode in a collection of freely hinging rigid elements~\cite{coulais2018characteristic}.
Such metamaterials display a plethora of exotic properties, such as tunable energy absorption~\cite{fan2022multistable}, programmability~\cite{florijn2014programmable, silverberg2014using, medina2020navigating, mueller2022architected}, self-folding~\cite{coulais2018multi, dieleman2020jigsaw}, nontrivial topology~\cite{kane2014topological, paulose2015topological, meeussen2020topological, ghatak2020observation} and shape-morphing~\cite{mullin2007pattern, shim2012buckling, cho2014engineering, waitukaitis2015origami, silverberg2015origami, dudte2016programming, overvelde2016three, overvelde2017rational, nevzerka2018jigsaw, dovskavr2023wang}. For shape-morphing in particular, a combinatorial framework was developed, where a small set of building blocks are tiled to form a
metamaterial~\cite{coulais2016combinatorial}. In all these examples, both the building blocks and the 
underlying mechanism exhibit a single zero mode, so that the metamaterial's response is dominated by a single soft mode leading to a single mechanical functionality. 
Often, by fixing the overall amplitude of deformation, the combinatorial design problem can be mapped to a spin-ice model~\cite{coulais2016combinatorial, meeussen2020topological, pisanty2021putting} or, similarly, to Wang tilings~\cite{nevzerka2018jigsaw, dieleman2020jigsaw, dovskavr2023wang}.

In contrast, multimodal metamaterials can potentially exhibit multiple functionalities~\cite{bossart2021oligomodal}.
Such metamaterials host multiple complex soft modes with potentially distinct functionalities. By controlling which mode is actuated, one can tune the metamaterial's response at will. 
To engineer such multimodal materials, one requires precise control over the structure and enumeration of zero modes.
However, 
as opposed to metamaterials based on building blocks with a single zero mode, 
the kinematics of multimodal metamaterials can no longer be captured by spin-ice or tiling problems. This is because 
linear combinations of zero modes are also valid zero modes 
such that the amplitudes of different deformation modes can take arbitrary values---such a problem can no longer be trivially mapped to a discrete tiling or spin-ice model.
As a consequence, 
designing multimodal materials is hard. Current
examples of multimodal metamaterials include those with tunable elasticity tensor and wave-function programmability~\cite{hu2023engineering}, and tunable nonlocal elastic resonances~\cite{bossart2023extreme}. In both works, the authors consider periodic lattices that limit the kinematic constraints between bimodal unit cells to (appropriate) boundary conditions, thereby allowing for straightforward optimization.
In contrast, we aim to construct design rules for 
{\em aperiodic} multimode structures that contain a large number of simpler bimodal building blocks and that exhibit a large, but controllable number of spatially aperiodic zero modes.
Such aperiodic modes allow for complex mechanical functionalities such as a strain-rate selectable auxetic response~\cite{bossart2021oligomodal} and sequential energy-absorption while retaining the original stiffness~\cite{liu2023leveraging}.
For aperiodic multimode structures,
the number of 
kinematic constraints grows with the size of the structure, so that successful designs require a large number of degeneracies between constraints.
A general framework to design such zero modes is lacking.

Here, we set a first step towards such a general framework for multimodal combinatorial metamaterials. We use this framework to find emergent combinatorial tiling rules for a multimodal metamaterial based on symmetries and degenerate kinematic constraints. Strikingly, we find nonlocal rules that restrict the type of tiles that are allowed to appear together \emph{anywhere} in the configuration. This is distinct from local tiling rules found in single-modal metamaterials which consist only of local constraints on pairs of tiles. 
Our work thus provides a new avenue for systematic design of spatial complexity, kinematic compatibility and multi-functionality in multimodal mechanical metamaterials.

To develop our framework, we focus on a recently introduced multimodal combinatorial metamaterial~\cite{bossart2021oligomodal}.
This metamaterial can host multiple complex zero modes that can be utilized to engineer functional materials. For example, a configuration of this metamaterial dressed with viscoelastic hinges allows for a strain-rate selectable auxetic response under uniaxial compression~\cite{bossart2021oligomodal}. Another recent example utilizes so-called \textit{strip modes} to efficiently absorb energy through buckling while retaining the original stiffness under sequential uniaxial compression~\cite{liu2023leveraging}. However, the design space remains relatively unexplored and is sufficiently rich and complex that further study of this combinatorial metamaterial is warranted.

\begin{figure}[t]
    \centering    \includegraphics{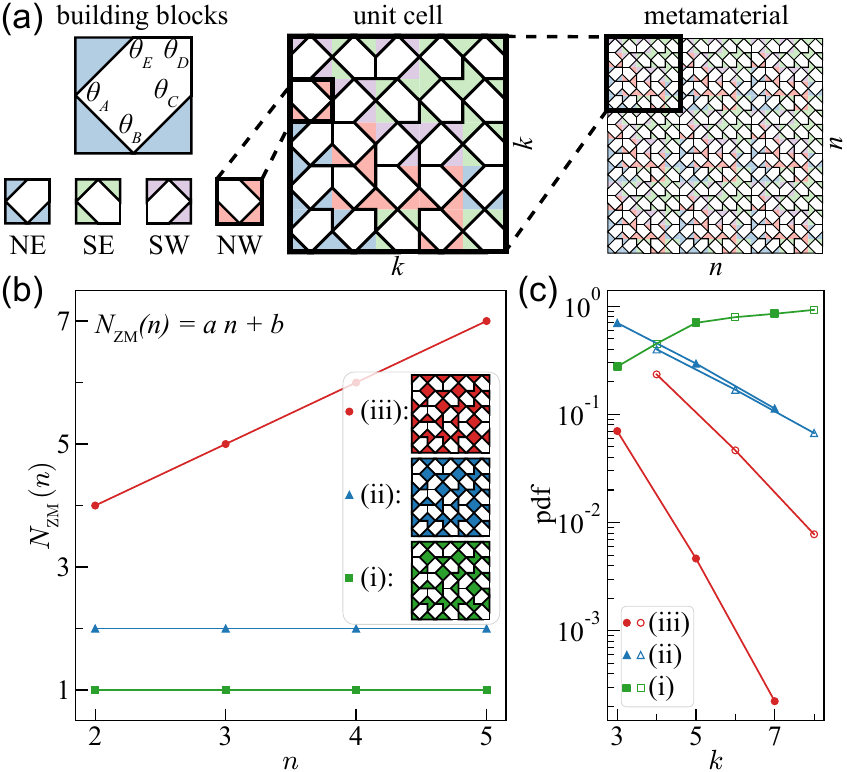}
    \caption{(a) Four differently oriented two-dimensional building blocks (left), combine into a square $k=5$ unit cell (middle) which is tiled in a $n=3$ grid to form a combinatorial metamaterial (right). The four orientations of the building block each have a unique color to guide the eye. The black lines represent rigid bars that hinge freely at intersections with other rigid bars. Colored regions are rigid polygons. \ch{We note that rigid pentagons with a reentrant edge are kinematically equivalent to rigid diamonds (rotated squares).} (b) The number of zero modes \ch{$N_{\mathrm{ZM}} (n)$} as a function $n$. We distinguish between three design classes, exemplified by the three unit cells designs shown in the legend. Note that the unit cells differ only by the rotation of a single building block, yet each belongs to another class. (c) Probability density function (pdf) to find each design class through Monte Carlo sampling of the design space. Class (ii) (blue \ch{triangles}) and (iii) (red \ch{circles}) become exponentially more rare with increasing unit cell size $k$, while class (i) (green \ch{squares}) becomes abundant~\cite{footnote_Uscaling}. The rate of exponential decline for class (ii) and (iii) depends on if $k$ is odd (filled) or even (open).}
    \label{fig:phenomenology}
\end{figure}

More concretely, this combinatorial metamaterial is
composed of building blocks consisting of rigid bars and hinges that feature two zero modes: deformations that do not stretch any of the bars to second order of deformation~\cite{bossart2021oligomodal} [Fig.~\ref{fig:phenomenology}(a)].
These degrees of freedom are restricted by kinematic constraints between neighboring building blocks, which in turn depend on how the blocks are tiled together. 
We stack these building blocks to form square $k\times k$ unit cells, and tile these
periodically to form metamaterials of $n \times n$ unit cells. 
These metamaterials can be classified in three distinct classes based on the number of zero modes \ch{$N_{\mathrm{ZM}}$} as function of $n$: most random configurations are
monomodal, due to the presence of a trivial global (counter-rotating)
single zero mode~\cite{bossart2021oligomodal, mastrigt2022machine}.
However, rarer configurations can be oligomodal (constant number $>1$ of zero modes) or plurimodal (number of zero modes proportional to $n$) [Fig.~\ref{fig:phenomenology}(b)].

The design space of this metamaterial was fully explored for $2\times 2$ and $3 \times 3$ unit cell tilings of such building blocks. For larger tilings, a combination of brute-force calculation of the zero modes and machine learning was used to classify the design space of larger unit cells up to $8\times 8$~\cite{mastrigt2022machine}. However, it is an open question how to construct design rules to determine this classification directly from the unit cell tiling without requiring costly matrix diagonalizations or machine learning. 

In this paper, we focus on the specific question of obtaining tiling rules for plurimodal designs for the aforementioned building blocks. Such plurimodes drive the mechanism behind the sequential energy-absorption metamaterial~\cite{liu2023leveraging}. A crucial role is played by degeneracies of the kinematic constraints. These kinematic constraints 
follow trivially from the tiling geometry and take the form of constraints between the deformation amplitudes of adjacent building blocks. For random tilings, the kinematic constraints rapidly proliferate, leading to the single trivial mode.
Checking for degeneracies between these constraints is nontrivial, as they are expressed as relations between the deformation amplitudes of different groups of building blocks. To check for degeneracies, we use a transfer matrix-like approach to map all these constraints to constraints on a small, pre-selected set, of deformation amplitudes. This allows us to establish a set of combinatorial rules. Strikingly, these  combine local tiling constraints on pairs of building blocks with 
global constraints on the types of tiles that are allowed to appear together;
hence, local information is not sufficient to identify a valid plurimodal tiling.

The structure of this paper is as follows. In Sec.~\ref{sec: phenomenology}
we investigate the phenomenology of this metamaterial, focusing on the number of zero modes \ch{$N_{\mathrm{ZM}}(n)$} for unit cell sizes $3\leq k \leq 8$. We show that random configurations are exponentially less likely to be oligomodal or plurimodal with increasing unit cell size $k$. Additionally, we define a mathematical representation of the building blocks' deformations that allows us to compare deformations in collections of building blocks.
In Sec.~\ref{sec: compatibility constraints} we derive a set of compatibility constraints on building block deformations that capture kinematic constraints between blocks. 
In Sec.~\ref{sec: mode structure} we use these constraints to formulate an exclusion rule that prohibits the structure of zero modes in collections of building blocks. Subsequently, we categorize the ``allowed'' mode-structures in three categories. 
In Sec.~\ref{sec: strip-modes} we devise a mode-structure that, if supported in a unit cell, should result in a linearly growing number of zero modes, i.e., the unit cell will be plurimodal. We define a set of additional constraints on deformations localized in a strip in the unit cell that should be satisfied to support a mode with such a mode-structure. We refer to such modes as `strip'-modes.
In section \ref{sec: transfer mapping formalism} we define a transfer matrix-like formalism that maps deformation amplitudes from a column of building blocks to adjacent columns.
In Sec.~\ref{sec: constraints and symmetries} we define a general framework using the transfer mappings defined in the previous section to determine if a strip of building blocks supports a strip mode of a given width $W$.
In Sec.~\ref{sec: deriving rules for strip-modes} we apply this framework explicitly on strips of width $1\leq W \leq 3$ and derive a set of tiling rules for strips of each width $W$. Surprisingly, we find that strips of width $W=3$ require a global constraint on the types of tiles that are allowed to appear together in the strip. Finally, we conjecture that there is a set of general design rules for strips of arbitrary width $W$, provide numerical proof of their validity and use them to construct a strip mode of width $W=10$.

\section{Phenomenology \label{sec: phenomenology}}
\textit{Configuration.}---We consider a family of hierarchically constructed  combinatorial metamaterials [Fig.~\ref{fig:phenomenology}(a)] \cite{bossart2021oligomodal}.
A single building block consist of three rigid triangles and two rigid bars that are flexibly linked, and its  deformations can be specified by the five interior angles $\theta_A, \theta_B, \dots, \theta_E$ that characterize the five hinges [Fig.~\ref{fig:phenomenology}(a)]. Each building blocks features two, linearly independent, zero energy deformations~\cite{bossart2021oligomodal, mastrigt2022machine}. As the undeformed building block has an outer square shape and inner pentagon shape, each building block can be oriented in four different orientations: $c=\{NE, SE, SW, NW\}$ [Fig.~\ref{fig:phenomenology}(a)]. We stack these building blocks to form square $k \times k$ unit cells. Identical unit cells are then periodically tiled to form metamaterials consisting of $n\times n$ unit cells; we use open boundary conditions. Each metamaterial is thus
specified by the value of $n$ and the design of the unit cell, given by the $k\times k$ set of orientations $C$.

\textit{Three classes.}---We focus on the number of zero modes \ch{$N_{\mathrm{ZM}}(n)$} (deformations that do not cost energy up to quadratic order) for a given design. In earlier work, we showed
that the number of zero modes is a linear function of $n$:
\ch{$N_{\mathrm{ZM}}  = a n + b$}, where $a\ge 0$ and $b\ge 1$ (see Fig.~\ref{fig:phenomenology}(b))~\cite{mastrigt2022machine}. Based on the values of $a$ and $b$, we define three design classes:
Class (i): $a = 0$ and $b=1$. For these designs, which become overwhelmingly likely for large $k$ random unit cells [Fig.~\ref{fig:phenomenology}(c)], there is a single global zero mode, which we will show to be the well known counter-rotating squares (CRS) mode~\cite{resch1965geometrical, grima2000auxetic, mullin2007pattern, bertoldi2010negative, shim2012buckling, coulais2015discontinuous, bertoldi2017flexible, coulais2018characteristic, coulais2018multi, czajkowski2022conformal};
Class (ii): $a=0$ and $b \geq 2$. For these rare designs, the metamaterial hosts additional zero modes that typically span the full structure, but \ch{$N_{\mathrm{ZM}}(n)$} does not grow with $n$;
Class (iii): $a \geq 1$. For these designs the number of zero modes grows linearly with system size $n$, and we will show that these rare zero modes are organized along strips.
Designs in
class (ii) and (iii) become increasingly rare with increasing unit cell size $k$ (see Fig.~\ref{fig:phenomenology}(c)). Yet, multi-functional behavior of the metamaterial requires
the unit cell design to belong to class (ii) or (iii). Hence we aim to find design rules that
allow to establish the class of a unit cell based on its real space configuration $C$ and that do not require costly diagonalizations to determine \ch{$N_{\mathrm{ZM}}(n)$}.
Such rules will also play a role for the designs of the rare configurations in class (ii) and (iii).

As we will show, deriving such rules requires a different analytical approach than previously used to derive design rules in mechanical metamaterials~\cite{coulais2016combinatorial, meeussen2020topological, pisanty2021putting, dieleman2020jigsaw} 
The reason is for this is that
each building block has two degrees of freedom yet potentially more than two nondegenerate constraints to satisfy. 
The problem can therefore not be mapped to a tiling problem~\cite{waitukaitis2015origami, dieleman2020jigsaw}.
In what follows, we will define an analytic framework based on transfer-mappings and constraint-counting and use this framework to derive design rules for unit cells of class (iii).

\begin{figure}[h]
    \centering
    \includegraphics{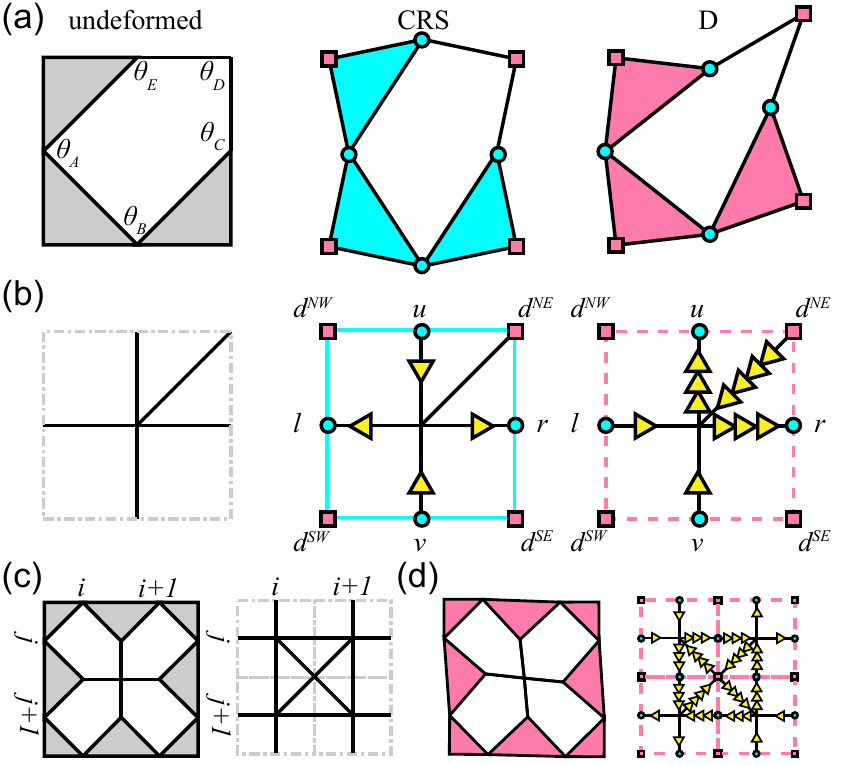}
    \caption{(a) Zero modes of the building block in orientation NE are infinitesimal deformations of the undeformed building block (left) expressed in the two basis zero modes CRS (middle) and D (right). These deformations are characterized by changes in the four angles on the faces of the block (cyan circles) and the four angles on the corners of the block (pink squares). 
    (b) The five interior angles of the building block in orientation NE are represented by edges in the bond representation (left). We express deformations of the building block as values on these edges, which we represented as arrows. The number of arrows corresponds to the magnitude of deformation, and the direction of the arrows (incoming, outgoing) to the sign. Note that the CRS mode (middle) deforms only the angles on the faces of the building block and thus does not depend on the orientation of the building block. However, the D mode (right) does deform a diagonal edge and the mode thus depends on the orientation of the building block.
    (c) Building blocks are tiled together on a grid to form unit cells (left, for a $2\times 2$ example), where the row index $j$ increases from top to bottom and the column index $i$ from left to right. The bond representation (right) forms the static background.
    (d) The static background is dressed with arrows on its bonds that represent deformations of the unit cell (left) in the vertex representation (right).
    }
    \label{fig: building block deformations}
\end{figure}

\textit{Zero modes of building blocks.}---To understand the spatial structure of zero modes, we first consider the zero energy deformations of an individual building block, irrespective of its orientation 
[Fig.~\ref{fig: building block deformations}(a)]. We can specify a zero mode $m_z$ of a single building block in terms of the infinitesimal deformations of the angles $\theta_A, \theta_B, \dots, \theta_E$, which we denote as
$\mathrm{d}\theta_A, \mathrm{d}\theta_B, \dots, \mathrm{d}\theta_E$, with respect to the undeformed, square configuration [Fig.~\ref{fig: building block deformations}(a)].
As the unit cell can be seen as a dressed five-bar linkage, it has two independent zero modes~\cite{bossart2021oligomodal, mastrigt2022machine}. We choose a basis where one of the basis vectors correspond to the Counter-Rotating Squares (CRS) mode, where $(\mathrm{d}\theta_A, \mathrm{d}\theta_B, \mathrm{d}\theta_C, \mathrm{d}\theta_D, \mathrm{d}\theta_E) \propto (1,-1,1,0,-1)$, and the other basis vector corresponds to what we call a `diagonal' (D) mode, where
$(\mathrm{d}\theta_A, \mathrm{d}\theta_B, \mathrm{d}\theta_C, \mathrm{d}\theta_D, \mathrm{d}\theta_E)  \propto (-1,-1,3,-4,3)$ [Fig.~\ref{fig: building block deformations}(a)]. A general deformation can then be written as
$(\mathrm{d}\theta_A, \mathrm{d}\theta_B, \mathrm{d}\theta_C, \mathrm{d}\theta_D, \mathrm{d}\theta_E) = \alpha (1,-1,1,0,-1) + \beta (-1,-1,3,-4,3)$, where $\alpha$ and $\beta$ are the amplitudes of the CRS-mode and D-mode, respectively.

\textit{Zero modes of unit cells.}---We now consider the deformations of a single building block in a fixed orientation.
Hence, we can express a zero mode of an individual building block $m_z$ as $m_z(\alpha_z, \beta_z, c_z) = \alpha_z m_{CRS} + \beta_z m_{D}(c_z)$. The deformation of each building block is completely determined by three degrees of freedom: the orientation $c_z$ and the amplitudes $\alpha_z$ and $\beta_z$ of the CRS and D mode. To compare these deformations for groups of building blocks, we now define additional notation.
We use a vertex representation~\cite{bossart2021oligomodal} where we map the changes in angles of the faces of the building block, $\mathrm{d}\theta_A, \mathrm{d}\theta_B, \mathrm{d}\theta_C$ and $\mathrm{d}\theta_E$ to values on horizontal ($l,r$) and vertical ($u,v$) edges, and the change in angle of the corner of the building block, $\mathrm{d}\theta_D$, to the value $d^c$ on a diagonal edge---note that the location of the diagonal edge represents the orientation, $c$, of each building block [Fig.~\ref{fig: building block deformations}(b)]. 
Irrespective of the orientation, we then find that a CRS mode corresponds to
$(u, v, l, r, d^{\mathrm{NE}}, d^{\mathrm{SE}}, d^{\mathrm{SW}}, d^{\mathrm{NW}}) \propto (-1, -1, 1, 1, 0, 0, 0, 0) = m_{CRS}$ [Fig.~\ref{fig: building block deformations}(b)]. For a D mode, the deformation depends on the orientation; for a NE block we have $(u, v, l, r, d^{\mathrm{NE}}, d^{\mathrm{SE}}, d^{\mathrm{SW}}, d^{\mathrm{NW}}) \propto (3, -1, -1, 3, -4, 0, 0, 0) = m_{D}(\mathrm{NE})$ [Fig.~\ref{fig: building block deformations}(b)].
We note that for a D mode in a building block with orientation $c$,
only a single diagonal edge is nonzero.
For ease of notation, we express the deformation of a building block with orientation $c$ in shorthand $(u, v, l, r, d^c)$, where the excluded diagonals are implied to be zero. In this notation, the D mode for a SE block is $(u, v, l, r, d^{\mathrm{SE}}) \propto (-1, 3, -1, 3, -4)=m_{D}(\mathrm{SE})$, for a SW block it is $(u, v, l, r, d^{\mathrm{SW}}) \propto (-1, 3, 3, -1, -4)=m_{D}(\mathrm{SW})$, and for a NW block it is $(u, v, l, r, d^{\mathrm{NW}}) \propto (3, -1, 3, -1, -4)=m_{D}(\mathrm{NW})$.
In addition, throughout this paper we will occasionally switch to a more convenient mode basis for calculation, where the degrees of freedom of $m_z$ are the orientation $c_z$ and the deformations $u_z$ and $v_z$.

To describe the spatial structure of zero mode deformations in a $k\times k$ unit cell, we place the building blocks on a grid and label their location as $(i, j)$, where the column index $i$ increases from left to right and the row index $j$ increases from top to bottom [Fig.~\ref{fig: building block deformations}(c)]. We label collections of the building block zero modes $m_{i, j}(\alpha_{i, j}, \beta_{i, j}, c_{i, j})$ as $M(A, B, C)$, where $A$, $B$, and $C$ are the collections of $\alpha_{i, j}$, $\beta_{i, j}$ and $c_{i, j}$. Such a collection $M(A, B, C)$ describes a \emph{valid} zero mode of the collection of building blocks $C$ if $M$'s elements, building block zero modes $m_{i, j}$, deform compatibly with its neighbors.

\section{Compatibility constraints \label{sec: compatibility constraints}}
Here, we aim to derive compatibility constraints on the deformations of individual building blocks in a collection of building block $C$ to yield a valid zero mode $M$ (Sec.~\ref{sec: phenomenology}). We find three local constraints that restrict the spatial structure of such valid zero modes.
First, we require compatible deformations along the faces between adjacent building blocks, and thus consider horizontal pairs (e.g., a building block at site $(i, j)$ with neighboring building block to its right at site $(i+1, j)$) and vertical pairs (e.g., a building block at site $(i, j)$ with neighboring building block below at site $(i, j+1)$)[Fig.~\ref{fig: building block deformations}(c)]. To be geometrically compatible,
the deformations of the joint face needs to be equal, yielding
\begin{equation}
    r_{i, j} = -l_{i+1, j}, \quad \mathrm{and} \quad v_{i, j} = -u_{i, j+1}
    \label{eq: compatibility constraints}
\end{equation}
for the `horizontal' and `vertical' compatibility constraints respectively.
Due to the periodic tiling of the unit cells, we need to take appropriate periodic boundary conditions into account; the deformations at faces located on the open boundary of the metamaterial are unconstrained.

Second, we require the deformations at the shared corners of four building blocks to be compatible. This yields the diagonal compatibility constraint [Fig.~\ref{fig: building block deformations}(c)]:
\begin{equation}
    d^{\mathrm{SE}}_{i, j} + d^{\mathrm{NE}}_{i, j+1} + d^{\mathrm{SW}}_{i+1, j} + d^{\mathrm{NW}}_{i+1, j+1}  = 0.
    \label{eq: plaquette constraint}
\end{equation}
We note that we again need to take appropriate periodic boundary conditions into account,
and note that the deformations at corners located on the open boundary of the metamaterial are unconstrained (see Appendix~\ref{App: open boundary conditions}).
For compatible collective deformations in a configuration of building blocks, we require these constraints to be satisfied for all sites, with appropriate boundary conditions: either periodic or open.

\section{Mode structure \label{sec: mode structure}}
In this section we determine an important constraint on the spatial structure of the zero modes that follows from the compatibility constraints [Eqs.~\eqref{eq: compatibility constraints} and \eqref{eq: plaquette constraint}]. 
We use the compatibility constraints to derive a constraint on the mode-structure of $2\times 2$ configurations, which in turn restricts the ``allowed'' spatial structures of valid zero modes $M$ in any configuration $C$.
To derive this constraint, we label the deformations of each building block as either CRS or D, depending on the magnitude of the D mode, $\beta_{i, j}$. We refer to building blocks with $\beta_{i, j} = 0$ as CRS blocks that deform as $m_{i, j} \propto m_{CRS}$, and to building blocks with $\beta_{i, j} \neq 0$ as D blocks. We will find that the compatibility constraints restrict the location of D and CRS blocks in zero modes.

Regardless of the unit cell configuration $C$, there is always a global CRS mode where all building blocks are of type CRS~\cite{bossart2021oligomodal, mastrigt2022machine}. To see this from our constraints, note that CRS blocks trivially satisfy the diagonal compatibility constraint [Eq.~\eqref{eq: plaquette constraint}], and when
we take $\alpha_{i, j} = (-1)^{i+j} \alpha$, also the horizontal and vertical compatibility constraints [Eq.~\eqref{eq: compatibility constraints}]. We refer to a deformation of CRS blocks that satisfies these constraints as an area of CRS with amplitude $\alpha$. Any configuration of building blocks with open boundaries supports a global area of CRS with arbitrary amplitude.
Another way to see this is to note that locally, the CRS mode $m_{CRS}$ does not depend on the building block's orientation $c$. 

\begin{figure}[t]
    \centering
    \includegraphics{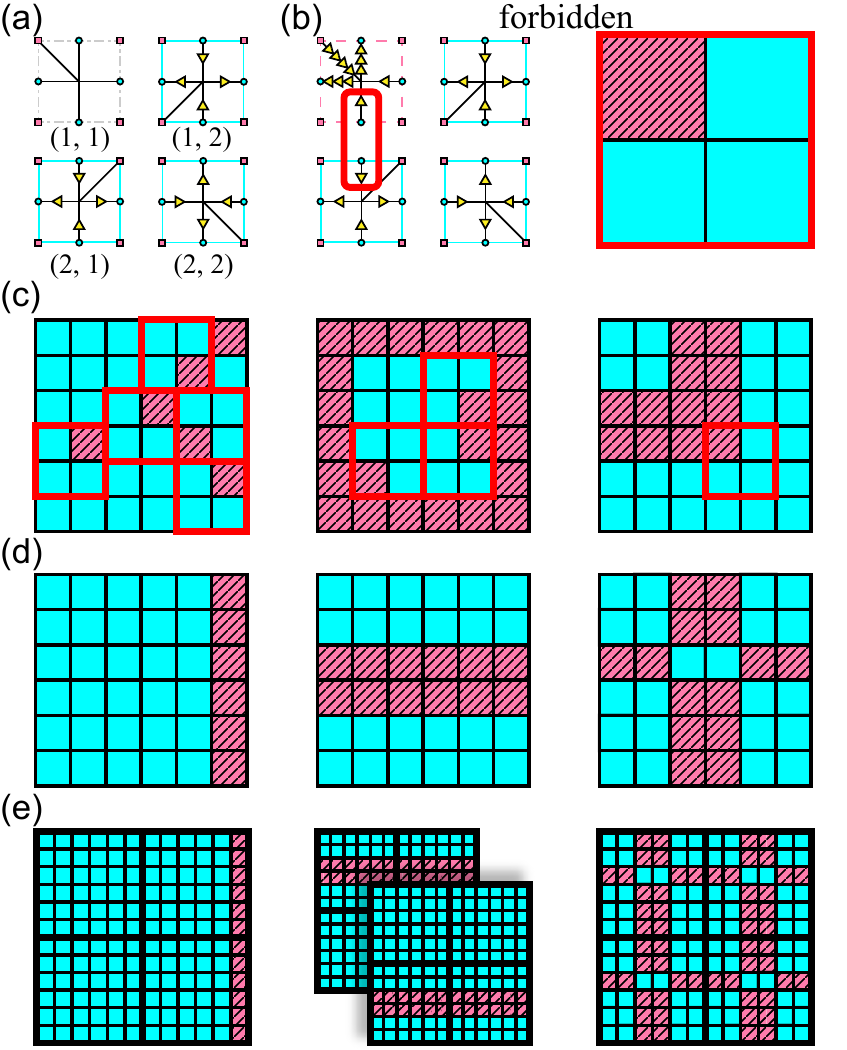}
    \caption{(a) $2\times 2$ configuration of building blocks with open boundaries. Three building blocks deform compatibly as CRS blocks (cyan \ch{solid} squares) with amplitude $\alpha=1$, while the top left building block is undetermined (gray \ch{dash-dotted} square). 
    (b) Left: example of an invalid zero mode. The top-left building block deforms incompatibly as a D block (pink \ch{dashed} square) with its CRS block neighbors (frustrated deformation is circled by \ch{thick} red square). Right: we describe the structure of a mode $M$ in CRS blocks \ch{(cyan and \ch{solid})} and D blocks \ch{(pink and striped)}. In general, a valid zero mode cannot contain any $2\times 2$ configurations that deform with a single D block surrounded by CRS blocks. Thus $2\times 2$ configurations with a single D block are forbidden, which we label by a thick red square. 
    (c) Forbidden zero mode structures for a $6\times 6$ configuration with open boundaries.
    (d) Allowed zero mode structures for a $6\times 6$ configuration with open boundaries. In Appendix~\ref{App: modeshape realizations} we show specific realizations of 'edge'-modes (left), 'stripe'-modes (middle), and 'Swiss cheese'-modes (right).
    (e) Zero mode structures for a $2\times 2$ tiling of a $6\times 6$ unit cell (thick black squares). 
    Note that the strip of D blocks in the stripe-mode (middle) can be located in both the top and bottom row of the tiling, and therefore leads to two valid zero modes in the tiling.
    }
    \label{fig:modeshapes}
\end{figure}

To find additional modes in a given configuration, at least one of the building blocks has to deform as type D. 
We now show that any valid zero mode in a $2\times 2$ plaquette cannot contain a single D block. 
Consider a $2 \times 2$ configuration of building blocks with an open boundary and assume that three of the building blocks deform as CRS blocks ($\beta_{1, 2}=\beta_{2,1} =\beta_{2,2} = 0$) [Fig.~\ref{fig:modeshapes}(a)].
These three blocks deform such that
\begin{equation}
    u_{2, 1} = -l_{1, 2}.
    \label{eq: 3 CRS}
\end{equation}
However, this is incompatible with a D block at site $(1,1)$---irrespective of its orientation, 
for a D block $v_{1, 1} \neq -r_{1, 1}$, so a D block is not compatible with 
three of such CRS blocks. Clearly, this argument does not depend on the specific location of the D blocks, 
since we are free to rotate the $2\times 2$ configuration and did not make any assumptions about the orientations of any of the building blocks. 
Hence, valid zero modes in any $2\times 2$ plaquette cannot feature a single D building block [Fig.~\ref{fig:modeshapes}(b)]. 

This implies that, first, in tilings that are at least of size $2\times 2$, D blocks cannot occur in isolation. Second, this implies that areas of CRS must always form a rectangular shape. To see this, consider zero modes with arbitrarily shaped CRS areas and consider $2\times 2$ plaquettes near its edge [Fig.~\ref{fig:modeshapes}(c)]. Any concave corner would locally feature a $2\times 2$ plaquette with a single D block, and is thus forbidden; only straight edges and convex corners are allowed. Hence, each area of CRS must be rectangular. 
In general, this means that in a valid zero mode the D and CRS blocks form a pattern of rectangular patches of CRS in a background of D [Fig.~\ref{fig:modeshapes}(d)]. 

Note that our considerations above only indicate which mode structures are forbidden. However, we have found that modes can take most ``allowed'' shapes, including `edge'-modes where the D blocks form a strip near the boundary, `stripe'-modes where the D blocks form system spanning strips, and `Swiss cheese'-modes, where a background of D blocks is speckled with rectangular areas of CRS [Fig.~\ref{fig:modeshapes}(d)].

We associate such modes with class (ii) or (iii) mode-scaling in unit cells. We observe that most edge-modes in a unit cell persist upon tiling of the unit cell by extending in the direction of the edge, resulting in a single larger edge-mode [Fig.~\ref{fig:modeshapes}(e)-left]. Swiss cheese-modes can also persist upon tiling of the unit cell by deforming compatibly with itself or another Swiss cheese-mode, creating a single larger Swiss cheese-mode [Fig.~\ref{fig:modeshapes}(e)-right]. Thus unit cells that support only edge-modes and Swiss cheese-modes have class (ii) mode-scaling. Moreover, we will show that a special type of stripe-mode, `strip'-modes, extend only along a single tiling direction, and allow for more strip modes by a translation symmetry [Fig.~\ref{fig:modeshapes}(e)-middle].
Here, we have found a rule on the deformations of $2\times 2$ plaquettes of building blocks that restricts the structure of valid zero modes in larger tilings.

\section{strip modes \label{sec: strip-modes}}
\begin{figure}[t]
    \centering
    \includegraphics{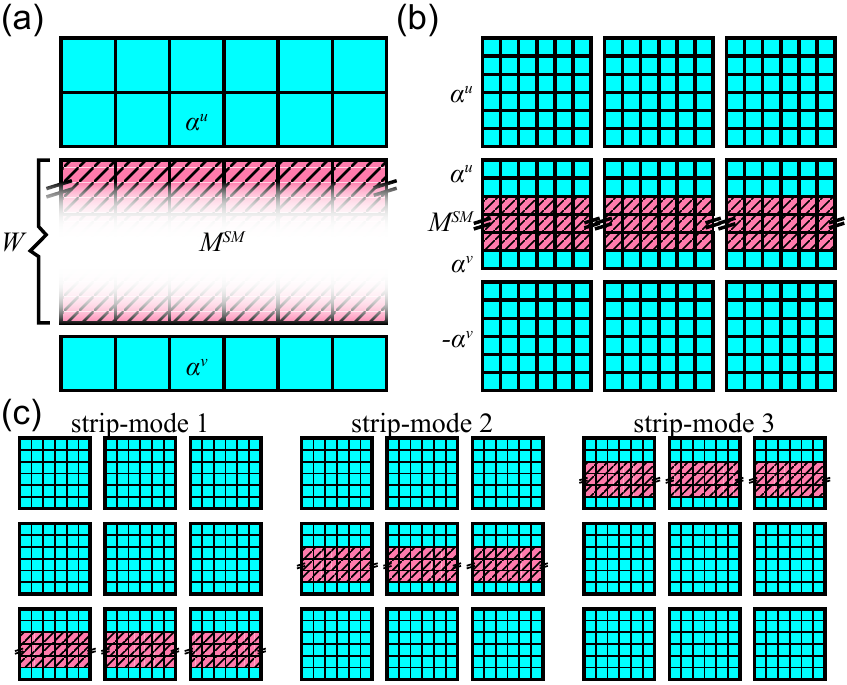}
    \caption{(a) Mode-structure of a strip mode in a $6\times 6$ unit cell. The strip of width $W$ deforms with strip deformation $M^{SM}$ (\ch{pink and} striped blocks) enclosed by two areas of CRS (\ch{cyan and solid} blocks) above and below the strip with CRS amplitudes $\alpha^u$ and $\alpha^v$.
    (b) Unit cells above and below the central unit cell deform compatibly with the strip mode as global areas of CRS. The sign of the CRS amplitudes depends on the parity of $k$ and the size of the area of CRS above and below the strip. Unit cells to the left and right of the central unit cell deform compatibly with the strip mode as strip modes.
    (c) A $6 \times 6$ unit cell with a $W=3$ strip that supports a strip mode is tiled to form a $3\times 3$ metamaterial. This metamaterial supports a strip mode in the bottom (left), middle (middle) and top (right) rows.}
    \label{fig: strip mode structure}
\end{figure}

We now focus on unit cells that are specifically of class (iii). We argue that a unit cell that can deform with the structure of a `strip'-mode is a sufficient condition for the number of modes \ch{$N_{\mathrm{ZM}}(n)$} to grow linearly with $a \geq 1$ for increasingly large $n\times n$ tilings. Here, we distinguish between stripe-modes and strip modes. We consider any zero mode that contains a deformation of non-CRS sites located in a strip enclosed by two areas of CRS a stripe-mode [Fig.~\ref{fig:modeshapes}(d)]. strip modes are a special case of stripe-modes: in addition to the aforementioned mode structure, we require the strip mode to deform compatibly (anti-)periodically across its lateral boundaries [Fig.~\ref{fig: strip mode structure}(a)]. As we will show, this requirement ensures that the strip mode persists in the metamaterial upon tiling of the unit cell and in turn leads to a growing number of zero modes with $n$. 
To find rules for unit cell configuration $C$ to support strip modes, we first in detail determine the required properties of strip modes for class (iii) mode-scaling. 
We then use these properties to impose additional conditions on the zero mode inside the strip of the configuration, strip conditions, and introduce a transfer matrix-based framework to find requirements on the configuration to support a strip mode.

We now consider the required properties of a strip mode for a $k\times k$ unit cell. We consider a unit cell in the center of a larger metamaterial that features a horizontal strip mode of width $W$ [Fig.~\ref{fig: strip mode structure}(a)]. In the strip mode, we take the areas outside the strip to deform as areas of CRS with amplitudes $\alpha=\alpha^u$ and $\alpha = \alpha^v$ for the areas above and below the strip respectively. We denote the deformation of the area inside the strip as $M^{SM}$ and require the strip to contain at least one D block. 
Compatibility between our central unit cell and its neighbors requires neighboring areas of CRS to be compatible. This is easy to do, as every unit cell is free to deform with a unit cell-spanning area of CRS. Thus the unit cells above and below the central unit cell deform compatibly with the strip mode if they deform completely as areas of CRS with equal or staggered CRS amplitude $\alpha^u$ and $\alpha^v$ [Fig.~\ref{fig: strip mode structure}(b)].
In addition, we require compatibility between the central unit cell and its left and right neighbors. Because the deformation in the strip $M^{SM}$ deforms compatibly with (anti-)periodic strip conditions across its lateral boundaries, unit cells to the right and left of the central unit cell deform compatibly with the strip mode if they deform as strip modes themselves [Fig.~\ref{fig: strip mode structure}(b)].
In an $n\times n$ tiling, all unit cells in any of the $n$ rows deforming as strip modes is a valid zero mode in the larger metamaterial [Fig.~\ref{fig: strip mode structure}(c)].
Therefore, we find a linearly increasing number of zero modes \ch{$N_{\mathrm{ZM}}(n)$} for unit cells that support a strip mode.

To find conditions on unit cell configurations $C$ to support a strip mode, we derive strip conditions from the structure of the strip mode on the strip deformation $M^{SM}$. Because areas of CRS are independent of the orientations of the building blocks in the area, we need only to find conditions on the configuration of building blocks in the strip $C^{SM}$. Without loss of generality, we focus on horizontal strip modes only. We consider a strip of building blocks $C^{SM}$ of length $k$ and width $W$ and relabel the indices of our lattice such that $(i, j) = (1, 1)$ corresponds to the upper-left building block in the strip: the row index is constrained to $1 \leq i \leq k$ and the column index is constrained to $1 \leq j \leq W$. For building blocks at the top of the strip to deform compatibly with an upper CRS area we require
\begin{equation}
    u_{i, 1} = -u_{i+1, 1},
    \label{eq: BC top}
\end{equation}
to hold along the entire strip.
We refer to this constraint as the upper strip condition. Without loss of generality we can set $u_{i, 1} = 0$ everywhere along the strip to ease computation, because we are free to add the global CRS mode with amplitude $-\alpha^u$ to the full strip mode so as to ensure that the upper deformation $u_{i, 1} = 0$ for all $i$.
Similarly, we require the building blocks at the bottom of the strip to satisfy
\begin{equation}
    v_{i, W} = -v_{i+1, W} 
    \label{eq: BC bottom}
\end{equation}
along the entire strip.
This constraint is referred to as the lower strip condition. Finally, we require the strip deformation to deform (anti-)periodically:
\begin{equation}
    \mathbf{v}_{1} =\begin{cases}
         (-1)^k \mathbf{v}_{k+1}, &\mathrm{if} \quad v_{1,W} \neq 0 \\
         |\mathbf{v}_{k+1}|, & \mathrm{if} \quad v_{1,W} = 0
    \end{cases}
    \label{eq: PBC}
\end{equation}
where the vector $\mathbf{v}_i = (v_{i, 1}, v_{i, 2}, ..., v_{i, W})$ fully describes the deformation of the building blocks in column $i$, if all deformations in the column satisfy the vertical compatibility constraints Eq.~\eqref{eq: compatibility constraints}. We refer to this condition as the periodic strip condition (PSC). 
We note that if the building blocks at the bottom of the strip deform as $v_{i, W} = 0$, both anti-periodic and periodic strip conditions result in a valid strip deformation.

Together with the horizontal and vertical compatibility constraints Eq.~\eqref{eq: compatibility constraints} and diagonal compatibility constraints Eq.~\eqref{eq: plaquette constraint}, the strip conditions Eq.~\eqref{eq: BC top} and Eq.~\eqref{eq: BC bottom} allow us to check if a configuration of building blocks in strip SM can satisfy all constraints and thus allow for a strip mode. 

\section{Transfer mapping formalism \label{sec: transfer mapping formalism}}
Now, we aim to derive necessary and sufficient requirements for configurations of building blocks in a strip of width $W$, $C^{SM}$, such that they allow for a valid strip deformation $M^{SM}$. 
To find such conditions, we introduce here transfer mappings that relate deformations in a column of building blocks to deformations in its neighboring columns. We will show later that these transfer mappings allow us to relate constraints and conditions on zero modes to requirements on the strip configuration.

\begin{figure}[t]
    \centering
    \includegraphics{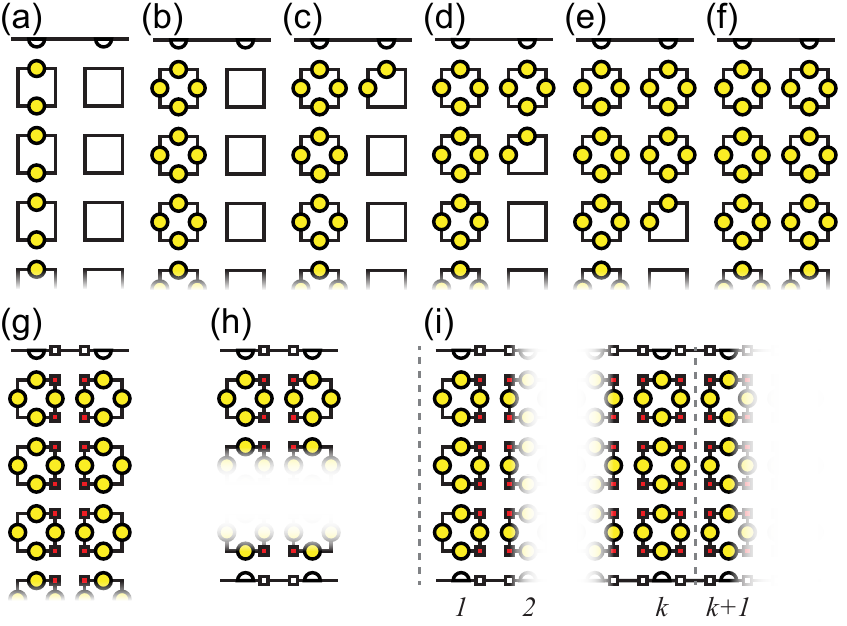}
    \caption{(a-f) Step-wise schematic illustration of our transfer mapping of deformations in a column of building blocks in the strip (white squares) to the next column in the strip, see main text. Yellow circles (\ch{light gray}) indicate known deformations of the building blocks, the upper white half-circles represent the upper strip condition [Eq.~\ref{eq: BC top}].
    (g-i) Schematic illustration of the constraints and conditions on the strip deformation, see main text. Red squares (\ch{dark gray}) indicate known diagonal deformation $d_z$ of the building blocks, the lower white half-circles represent the lower strip condition [Eq.~\eqref{eq: BC bottom}], and the lower numbers enumerate the columns for a strip of length $k$.}
    \label{fig:T schematic}
\end{figure}

To derive such transfer mappings, we first derive linear mappings between the pairs of degrees of freedom that characterize the zero mode $m_z$: the amplitudes of the CRS and D mode $(\alpha_z, \beta_z)$, the vertical edges $(u_z, v_z)$ and horizontal edges $(l_z, r_z)$. 
Subsequently, we derive a framework to construct strip modes: we fix the orientations $c_z$ throughout the strip ($C^{SM}$). We first fix the $(u_z, v_z)$ deformations for the left-most blocks in the strip [Fig.~\ref{fig:T schematic}(a)]. Then, using our linear maps, we determine $(l_z, r_z)$ for these blocks [Fig.~\ref{fig:T schematic}(b)]. We use the upper strip condition [Eq.~\ref{eq: BC top}] to determine
$u_z$ of the top block in the second column, and the horizontal compatibility constraint [Eq.~\ref{eq: compatibility constraints}] to determine $l_z$ of the second column [Fig.~~\ref{fig:T schematic}(c)]. Then we use a linear map to determine $(v_z)$ of the first block in the second column, and use vertical compatibility constraint [Eq.~\ref{eq: compatibility constraints}] to determine $(u_z)$ of the second block in the second column [Fig.~~\ref{fig:T schematic}(d)]. Repeating this last step, we obtain $(u_z, v_z)$ of the second column [Figs.~\ref{fig:T schematic}(e) and \ref{fig:T schematic}(f)], after which we can iterate this process to obtain  $(u_z, v_z, l_z, r_z)$ throughout the strip.
While above we have worked with upper and lower vertical edges $(u_z,v_z)$, we note that the deformations in a column follow from only the lower vertical edges $v_z$ in a column of building blocks $\mathbf{v}_i$, where $u_z$ follows from applying
the vertical compatibility constraint [Eq.~\eqref{eq: compatibility constraints}]. Thus, the deformation of building blocks in column $i+1$ is fully determined by the deformation in column $i$ by satisfying the vertical and horizontal compatibility constraints and the upper strip condition.

We refer to the linear mappings relating the deformations of column $i$, $\mathbf{v}_i$, to the deformations in adjacent column $i+1$, $\mathbf{v}_{i+1}$, as a linear transfer mapping $T(\mathbf{c}_{i}, \mathbf{c}_{i+1})$ which depends on the orientations of the building blocks in the two columns. Thus, by iterating this relation, the strip deformation is determined entirely by the deformations $\mathbf{v}_1$ of the left-most column.

\subsection{Linear degree of freedom transformations}
To derive these transfer mappings, we require linear mappings between the pairs of degrees of freedom that characterize the zero mode $m_z$.
For given set of orientations $\{c_z\}$, we derive linear mappings from the mode-amplitudes $(\alpha_z, \beta_z)$ to vertical edges
$(u_z, v_z)$ to horizontal edges $(l_z, r_z)$ and find that they all are nonsingular---this implies that any of these pairs fully characterizes the local soft mode $m_z$. 

First, we define $\Lambda$ as
\begin{eqnarray}
    \begin{pmatrix}
        u_z \\ v_z
    \end{pmatrix} &= \Lambda(c_z) \begin{pmatrix}
        \alpha_z \\ \beta_z
    \end{pmatrix}~.
\end{eqnarray}
Subsequently, 
we express $(l_z, r_z)$ in terms of $(u_z, v_z)$ as
\begin{eqnarray}
    \begin{pmatrix}
        l_z \\ r_z
    \end{pmatrix} &= \Gamma(c_z) \begin{pmatrix}
        \alpha_z \\ \beta_z
    \end{pmatrix} = \Gamma(c_z) \Lambda^{-1}(c_z) \begin{pmatrix}
        u_z \\ v_z
    \end{pmatrix}~.
\end{eqnarray}
Explicit expressions for the $2\times2$ matrices $\Lambda$ and $\Gamma$ are given in the Appendix~\ref{App: coordinate transform}.
Finally, we rewrite this equation as (see Table.~\ref{tab:L, R, D}):
\begin{eqnarray}
      \begin{pmatrix}
        l_z \\ r_z
    \end{pmatrix}
    =\begin{pmatrix}
        L_u(c_z) & L_v(c_z) \\ R_u(c_z) & R_v(c_z)
    \end{pmatrix} \begin{pmatrix}
        u_z \\ v_z
    \end{pmatrix}~.
    \label{eq: l, r in u, v}
\end{eqnarray}
Similarly, we can express the diagonal edge $d^o_z$ at orientation $o$ in terms of $(u_z, v_z)$ as (see Appendix~\ref{App: coordinate transform})
\begin{equation}
    d^o_z = D^o(c_z) (-u_z + v_z)~,
    \label{eq: d in u, v}
\end{equation}
where the coefficients $D^o(c_z)$ are given in Table~\ref{tab:L, R, D} for all orientations $o=\{ \mathrm{NE, SE, SW, NW} \}$.
We note that for CRS blocks where $u_z = v_z$ this equation immediately gives $d^o_z = 0$ for all orientations $o$. Together, Eqs.~\eqref{eq: l, r in u, v} and \eqref{eq: d in u, v} 
allow to express all building block deformations as linear combinations of the vertical deformations $(u_z, v_z)$.

\begin{table}[b]
    \centering
    \caption{
    Values for the coefficients $L_u, L_v, R_u, R_v$ for the $(u_z, v_z)$ to $(l_z, r_z)$ mapping [Eq.~\eqref{eq: l, r in u, v}] and the coefficient $D^o$ for the $(u_z, v_z)$ mapping to $d^o_z$ for a building block of orientation $c_z = \{\mathrm{NE, SE, SW, NW}\}$ [Eq.~\eqref{eq: d in u, v}].}
    \label{tab:L, R, D}
    \begin{ruledtabular}
    \begin{tabular}{l c  c  c  c}
                   &   NE  &   SE  &   SW  &   NW \\\hline
       $L_u$        &   -1/2    &   -1/2    &   -3/2    &   1/2    \\
        $L_v$       &    -1/2   &   -1/2    &   1/2    &    -3/2   \\
        $R_u$       &   1/2    &    -3/2   &    -1/2   &    -1/2   \\
        $R_v$       &   -3/2    &   1/2    &    -1/2   &    -1/2   \\
        $D^{\mathrm{NE}}$    &    1   &      0 &     0  &    0   \\
        $D^{\mathrm{SE}}$    &   0    &  -1     &    0   &   0    \\
        $D^{\mathrm{SW}}$    &   0    &  0     & -1      &   0    \\
        $D^{\mathrm{NW}}$    &   0    &   0    &   0    &   1  
    \end{tabular}
    \end{ruledtabular} 
\end{table}

\section{Constraints and Symmetries \label{sec: constraints and symmetries}}
Here, we define a general framework based on transfer-mappings and constraint-counting to determine if a given (strip) configuration $C^{SM}$ supports a valid strip mode $M^{SM}$.
The strip deformation $\mathbf{v}_1$ describes a valid strip mode only if it leads to a deformation which satisfies the diagonal compatibility constraints [Eq.~\ref{eq: plaquette constraint}] [Fig.~\ref{fig:T schematic}(g)], the lower strip conditions [Eq.~\ref{eq: BC bottom}] [Fig.~\ref{fig:T schematic}(h)] and the periodic strip condition [Eq.~\ref{eq: PBC}] [Fig.~\ref{fig:T schematic}(i)] everywhere along the strip. 
To determine if these constraints are satisfied by the deformation $\mathbf{v}_1$, we use the transfer mapping to map all the constraints throughout the strip to constraints on $\mathbf{v}_1$. 
Since each additional column yields additional constraints, we obtain a large set of constraints on $\mathbf{v}_1$, and without symmetries and degeneracies, one does not expect to find nontrivial deformations which satisfy all these constraints. However, for appropriately chosen orientations of the building blocks, many constraints are degenerate, due to the underlying symmetries.
Hence, we can now formulate two conditions for obtaining a nontrivial strip mode of width $W$. 

\begin{figure}[b]
    \centering
    \includegraphics{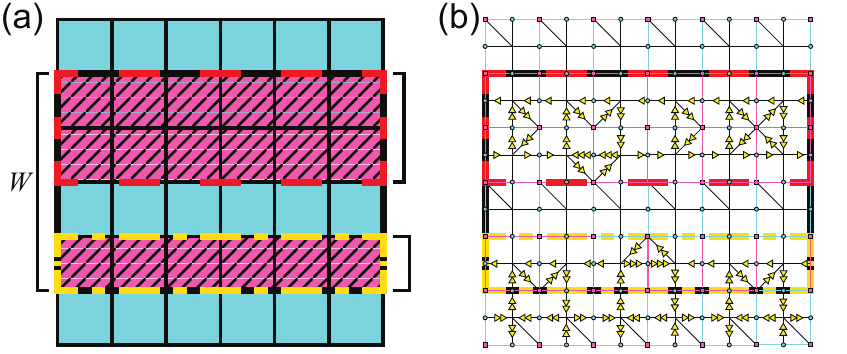}
    \caption{(a) A seemingly valid strip deformation of width $W=4$ (\ch{thick} black solid line) can be decomposed into two strips of smaller widths (\ch{thick, red dashed and yellow dash-dotted lines}) if it does not satisfy the CC and NT conditions. (b) Realization of a $W=4$ strip deformation (\ch{thick} black solid line) that does not satisfy the CC and NT conditions: it can be decomposed into $W'=2$ (enclosed in \ch{thick} red dashed line) and $W'=1$ (\ch{thick} yellow \ch{dash-dotted} line) strips that individually satisfy the NT and CC conditions.}
    \label{fig:IM}
\end{figure}

First, after mapping all the constraints in the strip to constraints on $\mathbf{v}_1$, and after removing redundant constraints, the number of nondegenerate constraints should equal $W-1$ so that the strip configuration contains a single non-CRS floppy mode. We refer to this condition as the constraint counting (CC) condition.
Second, we focus on irreducible strip modes of width $W$,
and exclude strip deformations composed of strip modes of smaller width or rows of CRS blocks [Fig.~\ref{fig:IM}(a)]. Such reducible strip deformations not only satisfy all constraints in a strip of width $W$, but also in an encompassing strip of width $W'<W$ [Fig.~\ref{fig:IM}(b)]. Irreducible strip modes of width $W$ do not satisfy all constraints for any encompassing strips of width $W'<W$.
We refer to this condition as the nontrivial (NT) condition as it excludes rows of CRS from the strip mode, which are trivial solutions to the imposed constraints. Valid strip modes are those that satisfy both CC and NT conditions.

To map all constraints to $\mathbf{v}_1$, we use the linear mapping between the diagonal edge $d_z$ and $(u_z, v_z)$ [Eq.~\eqref{eq: d in u, v}] such that the diagonal compatibility constraints [Eq.~\eqref{eq: plaquette constraint}] can be expressed in $v_z$. 
The diagonal compatibility constraints, lower strip conditions [Eq.~\eqref{eq: BC bottom}] and periodic strip condition [Eq.~\eqref{eq: PBC}] can all be expressed in $v_z$ and then be mapped to $\mathbf{v}_1$ by iteratively applying the set of transfer mappings $\{T(\mathbf{c}_i, \mathbf{c}_{i+1})\}$.

This constraint mapping method allows us to systematically determine if a given strip configuration $C^{SM}$ supports a valid strip mode $M^{SM}$:
\begin{enumerate}
    \item Determine the set of transfer matrices $\{T(\mathbf{c}_i, \mathbf{c}_{i+1}))\}$.
    \item Express the diagonal compatibility constraints [Eq.~\eqref{eq: plaquette constraint}], lower strip conditions [Eq.~\eqref{eq: BC bottom}] and periodic strip condition [Eq.~\eqref{eq: PBC}] in terms of $\{\mathbf{v}_i\}$.
    \item Map the set of all constraints to constraints on $\mathbf{v}_1$ using the transfer matrices.
    \item Check if the CC and NT conditions are satisfied on $\mathbf{v}_1$.
\end{enumerate}

In what follows, we consider the transfer mappings and constraints explicitly for strips of widths up to $W=3$ and derive geometric necessary and sufficient rules for the orientations $c_z$ of the building blocks to satisfy the CC and NT conditions.
Finally, we consider strips of even larger width $W$ and construct sufficient requirements on strip configurations. 

\section{Deriving rules for strip modes \label{sec: deriving rules for strip-modes}}
Here we aim to derive design rules for strip modes. We first derive necessary and sufficient conditions on strip configurations $C^{SM}$ of widths up to $W=3$. Then, we use those requirements to conjecture a set of general rules for strips of arbitrary widths. We provide numerical proof that these rules are correct and use them to generate a $W=10$ example that we would not have been able to find through Monte Carlo sampling of the design space.

\subsection{Case 1: $W=1$}
\begin{figure}[t]
    \centering
    \includegraphics{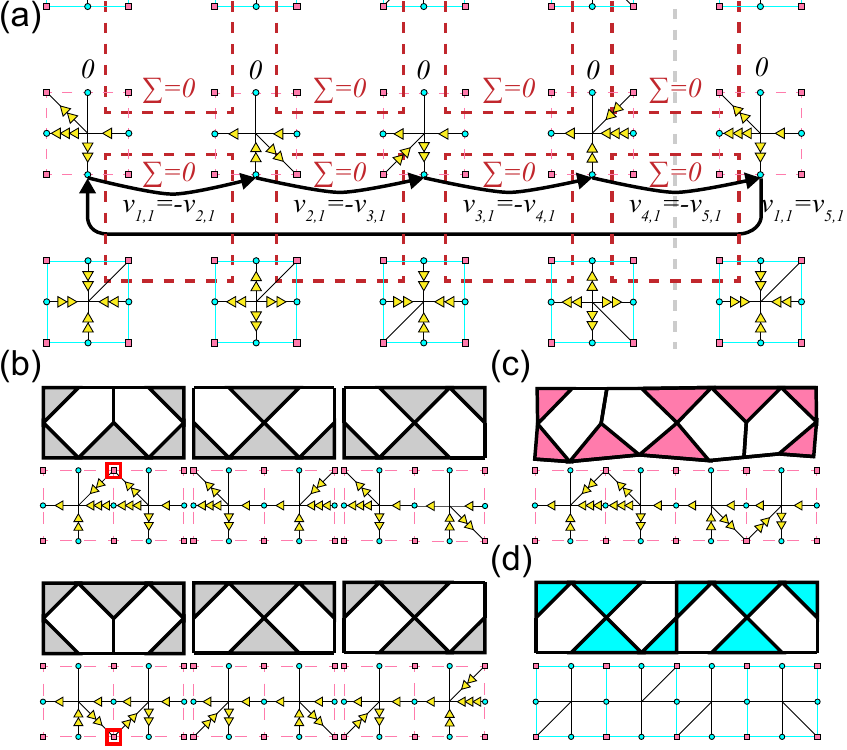}
    \caption{
    (a) Schematic representation of the degrees of freedom, constraints, and mapping for a $W=1$ strip mode of length $k=4$ in the vertex representation. 
    The building blocks in the strip and lower CRS area are highlighted with pink \ch{(dashed)} and blue \ch{(solid)} boxes; the upper CRS area has amplitude zero.
    Applying the horizontal compatibility constraint and upper strip condition leads to a mapping from $v_{i, 1}$ to $v_{i+1, 1}$. We show the deformation of each building block in the strip for such a mapping with $v_{1, 1} = 2$. The diagonal compatibility constraints are indicated by $\sum = 0$ in \ch{thick} red dashed boxes and are all satisfied by the mapping. The lower strip condition ($v_{i, 1} = -v_{i+1, 1}$, arrows) and periodic strip condition ($v_{1, 1} = v_{5, 1}$, long arrow) are also satisfied by the mapping. The strip therefore deforms compatibly with the lower CRS area with amplitude two.
    (b) The six h-pairs of horizontally adjacent building blocks $(c_{i, 1}, c_{i+1, 1})$ that satisfy Eq.~\eqref{eq: RLh} and examples of their deformations in vertex representation obtained from the map [Eq.~\eqref{eq: W1 map}] with $v_{i, 1} = 2$. Note that $d_{i, 1}^{NE} = - d_{i+1, 1}^{NW}$ and $d_{i, 1}^{SE} = - d_{i+1, 1}^{SW}$ are satisfied either trivially or by the transfer mapping [Eq.~\eqref{eq: W1 map}] (corner nodes highlighted with \ch{thick} red squares) for all h-pairs.
    (c) Example of a $k=4$ strip configuration (top) deformed as a valid strip mode $M^{SM}$ (bottom, vertex representation) that satisfies all compatibility constraints and strip conditions.
    (d) Example of a $k=4$ strip configuration (top) that can only satisfy all compatibility constraints and strip conditions by not deforming (bottom, vertex representation).
    } 
    \label{fig:W1constraints}
\end{figure}

We now derive necessary and sufficient conditions on the orientations of the building blocks
for strip modes of width $W=1$ to appear [Fig.~\ref{fig:W1constraints}(a)]. 
We show that a simple pairing rule for the orientations of neighboring building blocks gives necessary and sufficient conditions for such a configuration to support a valid strip mode, i.e., a strip deformation that satisfies the horizontal compatibility constraints [Eq.~\eqref{eq: compatibility constraints}], the diagonal compatibility constraints [Eq.~\eqref{eq: plaquette constraint}], the upper strip conditions [Eq.~\eqref{eq: BC top}], the lower strip conditions [Eq.~\eqref{eq: BC bottom}], and the periodic strip condition [Eq.~\eqref{eq: PBC}[] (see Fig.~\ref{fig:W1constraints}(a))
in addition to the constraint counting (CC) and nontrivial (NT) conditions.

First, we derive the transfer mapping that maps the deformations of building block
$(i,1)$ to block $(i+1,1)$ for general orientations $(c_{i, 1}, c_{i+1, 1})$. Without loss of generality, we set the amplitude $\alpha^u= 0$ such that $u_{i, 1} = 0$ everywhere along the strip---this trivially satisfies the upper strip condition [Eq.~\eqref{eq: BC top}] (recall that we can always do this by adding a global CRS deformation of appropriate amplitude to a given mode). The deformations of each building block are now completely determined by choosing $v_{i, 1}$. However, these cannot be chosen independently due to the various constraints. Implementing the horizontal compatibility constraints and upper strip condition, we find that the $v_{i, 1}$ in adjacent blocks are related via a linear mapping (see Appendix~\ref{App: transfer mapping}):
\begin{equation}
    v_{i+1, 1} = - \frac{R_v(c_{i, 1})}{L_v(c_{i+1, 1})} v_{i, 1}~,
    \label{eq: W1 map}
\end{equation}
where the values of $R_v(c)$ and $L_v(c)$ are given in Table~\ref{tab:L, R, D}.
We interpret this mapping as a simple (scalar) version of a transfer mapping (see Fig.~\ref{fig:W1constraints}(a)). The idea is then that, by choosing $v_{1, 1}$ and iterating the map [Eq.~\eqref{eq: W1 map}], we determine a strip deformation which satisfies both the upper strip conditions and horizontal compatibility constraints. The goal is to find values for the orientations $c_{i, 1}$ that produce a \emph{valid} strip mode, i.e., a deformation which also satisfies the diagonal compatibility constraints [Eq.~\eqref{eq: plaquette constraint}, red dashed boxes in Fig.~\ref{fig:W1constraints}(a)], lower strip conditions [Eq.~\eqref{eq: BC bottom}, black arrows in Fig.~\ref{fig:W1constraints}(a)], periodic strip condition [Eq.~\eqref{eq: PBC}, long black arrow in Fig.~\ref{fig:W1constraints}(a)], and CC and NT conditions---note that if we take $v_{1,1} =0$, all deformations throughout the unit cell are zero and we have simply obtained a zero amplitude CRS mode, which is not a valid strip mode [see example in Fig.~\ref{fig:W1constraints}(d)].

To construct configurations that produce a valid strip mode, we first consider an example. In this example, we only consider orientations $(c_{i, 1}, c_{i+1, 1})$ that satisfy
\begin{equation}
R_v(c_{i, 1}) = L_v(c_{i+1, 1})~,
\label{eq: RLh}
\end{equation}
and show that this is a sufficient 
condition to produce a valid strip mode. We refer to the six pairs $(c_{i, 1}, c_{i+1, 1})$ that satisfy condition Eq.~\eqref{eq: RLh}
as h-pairs (for horizontal) [Fig.~\ref{fig:W1constraints}(b)].

We find that configurations consisting only of h-pairs satisfy the lower and periodic strip conditions and diagonal compatibility constraints. Specifically, we find the following for h-pairs
\begin{enumerate}[(1)]
    \item  the map Eq.~\eqref{eq: W1 map} simplifies to 
    $v_{i+1, 1} = - v_{i, 1}$ and thus directly satisfies the lower strip condition [Eq.~\eqref{eq: BC bottom}] and periodic strip condition [Eq.~\eqref{eq: PBC}] by iterating the map, see deformations in Fig.~\ref{fig:W1constraints}(b).

    \item the diagonal compatibility constraints are either trivially satisfied or the same as the map Eq.~\eqref{eq: W1 map} and thus impose no constraints on $v_{i, 1}$. To see this, note that the diagonal compatibility constraint [Eq.~\eqref{eq: plaquette constraint}] is required to be satisfied at all corner nodes in the strip (pink squares in Fig.~\ref{fig:W1constraints}(a)). Note that away from the strip, all diagonals are zero (recall that a CRS block always has $d^c=0$). Thus, the diagonal compatibility constraint at the corner nodes shared between two building blocks in a pair simplifies to $d_{i, 1}^{NE} = d_{i+1, 1}^{NW}$ and $d_{i, 1}^{SE} = d_{i+1, 1}^{SW}$ (see Fig.~\ref{fig:W1constraints}(a)).
    For the six h-pairs, there are four pairs where all diagonals in the constraints are zero, i.e., trivially satisfied, and two pairs where the diagonals are nonzero (highlighted in red in Fig.~\ref{fig:W1constraints}(b)). For the latter case, the diagonal compatibility constraint implies that $v_{i, 1} = -v_{i+1, 1}$---this follows from $u_{i, 1} = 0$ and the mapping [Eq.~\eqref{eq: d in u, v}]---which is the same as the map Eq.~\eqref{eq: W1 map}.
\end{enumerate}
Thus, all conditions and constraints are trivially satisfied for strip configurations consisting only of h-pairs, see Fig.~\ref{fig:W1constraints}(c) for an example.

Such strip configurations thus impose no constraints on $\mathbf{v}_1$, thereby satisfying the constraint counting (CC) condition. Additionally, such configurations satisfy the nontrivial (NT) condition as well so long as $v_{1, 1} \neq 0$.
Hence, the pairing rule
\begin{enumerate}[(i)]
    \item Every pair of horizontally adjacent building blocks in the strip must be an h-pair. \label{rule: w1}
\end{enumerate}
is a sufficient condition to obtain valid $W=1$ strip modes, and thus class (iii) mode scaling.
It is also a necessary condition, because any pair that does not satisfy condition Eq.~\eqref{eq: RLh} does not trivially satisfy the lower strip condition [Eq.~\eqref{eq: BC bottom}], breaking the CC condition, and thus only satisfies all compatibility constraints and strip conditions of a strip mode for $v_{1, 1} = 0$, 
breaking the NT condition, see Fig.~\ref{fig:W1constraints}(d) for an example. Concretely, when $u_{1, 1}$ and $v_{1, 1}$ are both zero, the whole deformation is zero which is not a valid strip mode but rather a zero amplitude CRS mode. Hence, the pairing rule~\eqref{rule: w1} is a necessary and sufficient condition to obtain $W=1$ strip modes.

\subsection{Case 2: $W=2$}
Now, we consider strips of width $W=2$. strip deformations in such strips have an additional degree of freedom, $v_{i, 2}$, compared to strips of width $W=1$. To result in a valid strip mode there must be one constraint on the strip deformation $\mathbf{v}_1$ to satisfy the constraint counting (CC) condition. We show that a simple adjustment and addition to the pairing rule results in a sufficient and necessary condition to obtain $W=2$ strip modes.

\begin{figure*}[t]
    \centering
    \includegraphics{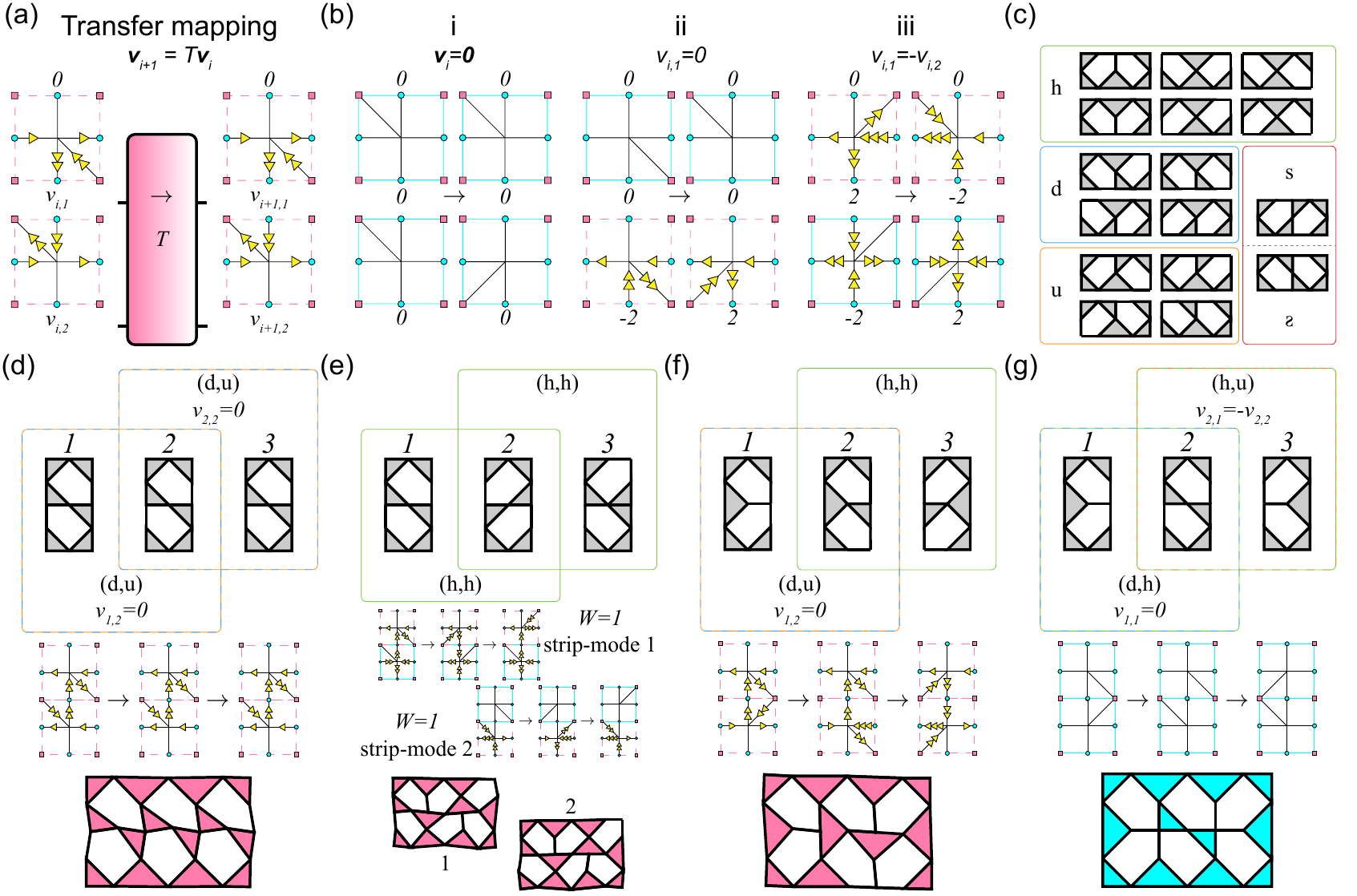}
    \caption{(a) The transfer matrix $T(\mathbf{c}_i, \mathbf{c}_{i+1})$ maps the displacements $\mathbf{v}_i = (v_{i, 1}, v_{i, 2})$ of the building blocks in column $i$, $\mathbf{c}_i$, to the displacements $\mathbf{v}_{i+1}$ of the building blocks in column $i+1$, $\mathbf{c}_{i+1}$, indicated by $\rightarrow$.
    (b) Three constraints on the strip deformation $\mathbf{v}_i$ that break the nontrivial (NT) condition in any $2\times 2$ strip configuration. The strip deformation can only satisfy the constraint by deforming rows of the strip as blocks of CRS (\ch{solid} cyan); either both rows (i), the top row (ii) or the bottom row (iii).
    (c) The 16 possible pairs of horizontally adjacent building blocks can be divided in four categories: horizontal (h, green), down (d, blue), up (u, orange), and vertical (s or \reverses, red). 
    (d) A strip configuration (top) consisting solely of (d, u)-pairs can deform as a valid $W=2$ strip mode. A realization of valid strip mode with $\mathbf{v}_1 = (-2, 0)$ is shown in vertex representation (middle) and schematic representation (bottom).
    (e) A strip configuration (top) consisting solely of (h, h)-pairs supports two $W=1$ strip modes (middle: vertex representation, bottom: schematic representation), as (h, h)-pairs impose no constraint on the strip deformation thereby breaking the constraint counting (CC) condition. Note that only part of the strip is shown; a strip consisting only of (h, h)-pairs must always have an even strip length $k$.
    (f) A strip configuration (top) consisting of a (d, u)-pair and (h, h)-pair. The (d, u)-pair imposes the constraint $v_{1, 2} = 0$ on the strip deformation $\mathbf{v}_1$. A realization of a valid strip mode with $\mathbf{v}_1=(-2, 0)$ is shown in vertex representation (middle) and schematic representation (bottom).
    (g) An invalid strip configuration (top), consisting of a (d, h)- and (h, u)-pair. The constraints imposed on the strip deformation, $v_{1, 1} = 0$ and $v_{2, 1} = -v_{2, 2}$, result in the strip being unable to deform, i.e., $\mathbf{v}_1 = (0, 0)$ (middle: vertex representation, bottom: schematic representation).
    }
    \label{fig: width 2}
\end{figure*}

First, we extend our transfer mapping to account for the extra row of building blocks in the strip. 
We again set the amplitude $\alpha^u=0$, so that the deformations of column $i$ are completely determined by fixing vector $\mathbf{v}_i = (v_{i, 1}, v_{i, 2})$ [Fig.~\ref{fig:T schematic}]. 
We now aim to obtain a complete map from $\mathbf{v}_{i}$ to $\mathbf{v}_{i+1}$.
Note that the map for $v_{i+1, 1}$ does not depend on the extra row of building blocks and therefore follows the map [Eq.~\eqref{eq: W1 map}] derived for $W=1$ strip modes. 
To obtain a map for $v_{i+1, 2}$, we note that for the building blocks in column $i+1$ to deform compatibly, we require the vertical compatibility constraint [Eq.~\eqref{eq: compatibility constraints}] to be satisfied [Fig.~\ref{fig:T schematic}]. 
Then, by implementing the horizontal and vertical compatibility constraints, we find a linear mapping for $v_{i+1, 2}$ which depends on both $v_{i, 1}$ and $v_{i, 2}$ (see Appendix~\ref{App: transfer mapping}):
\begin{eqnarray}
    v_{i+1, 2} &= \frac{L_u(c_{i+1, 2})}{L_v(c_{i+1, 2})} \left ( \frac{R_u(c_{i, 2})}{L_u(c_{i+1, 2})} - \frac{R_v(c_{i, 1})}{L_v(c_{i+1, 1})} \right) v_{i, 1} \nonumber\\
    & - \frac{R_v(c_{i, 2})}{L_v(c_{i+1, 2})} v_{i, 2}~.
    \label{eq: W2 map}
\end{eqnarray}
Together, Eq.~\eqref{eq: W1 map} and Eq.~\eqref{eq: W2 map} form the transfer mapping from $\mathbf{v}_i$ to $\mathbf{v}_{i+1}$, which we capture compactly as $\mathbf{v}_{i+1} = T(\mathbf{c}_i, \mathbf{c}_{i+1}) \mathbf{v}_i$ (see Fig.~\ref{fig: width 2}(a) for a schematic representation), where :
\begin{eqnarray}
    T(\mathbf{c}_i, \mathbf{c}_{i+1}) = \nonumber\\
    \begin{pmatrix}
        - \frac{R_v(c_{i, 1})}{L_v(c_{i+1, 1})} & 0\\
        \frac{L_u(c_{i+1, 2})}{L_v(c_{i+1, 2})} \left ( \frac{R_u(c_{i, 2})}{L_u(c_{i+1, 2})} - \frac{R_v(c_{i, 1})}{L_v(c_{i+1, 1})} \right) & -\frac{R_v(c_{i, 2})}{L_v(c_{i+1, 2})}
    \end{pmatrix}~.
    \label{eq: T W2}
\end{eqnarray}
Note that $T(\mathbf{c}_i, \mathbf{c}_{i+1})$ is a lower-triangular transfer matrix which depends only on the orientations $\mathbf{c}_i = (c_{i, 1}, c_{i, 2})$ of column $i$ and column $i+1$.

Now, we want to find values for the orientations $\mathbf{c}_i$ that produce a valid strip mode, i.e., a deformation which satisfies all constraints: the diagonal compatibility constraints [Eq.~\eqref{eq: plaquette constraint}], the lower strip condition [Eq.~\eqref{eq: BC bottom}] and periodic strip condition [Eq.~\eqref{eq: PBC}]. Additionally, the strip deformation $\mathbf{v}_1$ should satisfy the CC and NT conditions. We note that $\mathbf{v}_1 = \mathbf{0}$ corresponds to the strip deforming as an area of CRS [Fig.~\ref{fig: width 2}(b)-i]. Additionally, $v_{1, 1} = 0$ while $v_{1, 2}\neq 0$ corresponds to the top row deforming as an area of CRS with zero amplitude [Fig.~\ref{fig: width 2}(b)-ii] and $v_{1, 1} = - v_{1, 2}$ corresponds to the bottom row deforming as an area of CRS with arbitrary amplitude [Fig.~\ref{fig: width 2}(b)-iii, see Appendix~\ref{App: NT conditions}]. All these cases break the nontrivial (NT) condition as they describe strip deformations completely or in-part composed of rows of CRS blocks and thus do not represent valid $W=2$ strip modes. We exclude these configurations. 

To construct valid strip configurations, we consider $2\times 2$ configurations of building blocks $(\mathbf{c}_i, \mathbf{c}_{i+1})$. We compose such $2\times 2$ configurations by vertically stacking pairs of horizontally adjacent building blocks $(c_{i, 1}, c_{i+1, 1})$ and $(c_{i, 2}, c_{i+1, 2})$ for the top row and bottom row. 
There are 16 different pairs, and we note these can be grouped in four categories, depending on the corresponding values of $R_u, R_v, L_u$ and $L_v$ (Table~\ref{tab:L, R, D}):
\begin{eqnarray}
    \mathrm{h-pairs:} \quad & \frac{R_u(c_{i, j})}{L_u(c_{i+1, j})} = \frac{R_v(c_{i, j})}{L_v(c_{i+1, j})} = 1~, \label{eq: cond h} \\
    \mathrm{u-pairs:} \quad & \frac{R_u(c_{i, j})}{L_u(c_{i+1, j})} = -1~,\label{eq: cond u} \\
    \mathrm{d-pairs:} \quad & \frac{R_v(c_{i, j})}{L_v(c_{i+1, j})} = -1~, \label{eq: cond d}\\
    \mathrm{s-pairs:} \quad & \frac{R_u(c_{i, j})}{L_v(c_{i+1, j})} = \frac{R_v(c_{i, j})}{L_u(c_{i+1, j})} = 1~. \label{eq: cond s}
\end{eqnarray}
Each of the sixteen possible pairs satisfy only one of these conditions [Fig.~\ref{fig: width 2}(c)]. We denote groups of $2 \times 2$ configurations as vertical stacks of such pairs, e.g., a (d, u)-pair obeys the condition for d-pairs [Eq.~\eqref{eq: cond d}] for $(c_{i, 1}, c_{i+1, 1})$ and the condition for u-pairs [Eq.~\eqref{eq: cond u}] for $(c_{i, 2}, c_{i+1, 2})$; see Figs.~\ref{fig: width 2}(d) and \ref{fig: width 2}(f) for examples of (d, u)-pairs.

By stacking pairs, there are $16^{2}$ possible $2\times 2$ configurations. We now show that (d, u)-pairs and (h, h)-pairs are the only $2\times 2$ configurations that make up strip configurations that support valid $W=2$ strip modes. First, we will show that a strip composed only of (d, u)-pairs results in a valid strip mode. Second, we show that a strip composed only of (h, h)-pairs does not result in a single $W=2$ strip mode, but in two $W=1$ strip modes, breaking the CC condition. Finally, we show that combining (h, h)-pairs and (d, u)-pairs in a strip configuration results in a valid $W=2$ strip mode.

First, we consider (d, u)-pairs and show that these satisfy all conditions for a valid strip mode, provided that a single constraint on $\mathbf{v}_i$ is satisfied. First, from Eq.~\eqref{eq: cond u} and Eq.~\eqref{eq: cond d} we see that such pairs satisfy the condition
\begin{equation}
    \frac{R_u(c_{i, 2})}{L_u(c_{i+1, 2})} = \frac{R_v(c_{i, 1})}{L_v(c_{i+1, 1})}~,
    \label{eq: d u}
\end{equation}
which implies that the transfer matrix $T(\mathrm{(d, u)})$ [Eq.~\eqref{eq: T W2}] is purely diagonal. The map [Eq.~\eqref{eq: W2 map}] from $v_{i, 2}$ to $v_{i+1, 2}$ is thus independent of $v_{i, 1}$.
We now show that the choice $\mathbf{v}_1 = (v_{1, 1}, 0)$, which satisfies the constraint $v_{1, 2} = 0$, produces a valid strip mode for $v_{1, 1} \neq 0$, see Fig.~\ref{fig: width 2}(d) for an example strip deformation. This choice clearly satisfies the lower strip condition [Eq.~\eqref{eq: BC bottom}].
Moreover, the diagonal compatibility constraints [Eq.~\eqref{eq: plaquette constraint}] on corner nodes between the two columns $i$ and $i+1$ are also satisfied by the constraint $v_{i, 2} = 0$, regardless of the precise orientations of the building blocks as can be shown (see Appendix~\ref{App: Diag W2}).
Finally, by iterating the transfer map [Eq.~\eqref{eq: T W2}] for a strip that consists only of (d, u)-pairs, we find that $v_{1, 1} = v_{k+1, 1}$ and $v_{1, 2} = v_{k+1, 2} = 0$, i.e., the periodic strip condition [Eq.~\eqref{eq: PBC}] is satisfied.
Thus, a strip consisting only of (d, u)-pairs satisfies all constraints in the strip by imposing a single constraint on $\mathbf{v}_1$, satisfying the CC condition, and satisfies the NT condition so long as $v_{1, 1} \neq 0$. The resulting strip deformation is characterized by the choices of $c_{i, j}$ and $\mathbf{v}_1 = (v_{1, 1}, 0)$.

Second, we consider (h, h)-pairs and show that, while satisfying the diagonal compatibility constraints [Eq.~\eqref{eq: plaquette constraint}], lower strip conditions [Eq.~\eqref{eq: BC bottom}] and periodic strip conditions [Eq.~\eqref{eq: PBC}], they in fact lead to two adjacent $W=1$ strip modes, breaking the CC condition. 
Using Eq.~\eqref{eq: cond h} and the definition of the transfer matrix, we find that $T(\mathrm{(h, h)}) = -I$, where $I$ is the identity matrix.
Thus, (h, h)-pairs trivially satisfy the lower strip condition and diagonal compatibility constraints (see Appendix~\ref{App: lower strip condition}, see Fig.~\ref{fig: width 2}(e) for examples of strip deformations). Additionally, a strip that consists only of (h, h)-pairs maps $v_{1, j} = (-1)^k v_{k+1, j}$ by iterating the transfer mapping [Eq.~\eqref{eq: T W2}] and thus satisfies the periodic strip condition. However, a strip which consists only of (h, h)-pairs does not place any constraints on $\mathbf{v}_1$ and retains the two degrees of freedom that each can describe valid $W=1$ strip modes [Fig.~\ref{fig: width 2}(e)], breaking the CC condition. Thus, a strip composed only of (h, h)-pairs does not support one $W=2$ strip mode, but two $W=1$ strip modes. 

We now consider combining (h, h)-pairs and (d, u)-pairs in a single strip and show that such a strip supports a valid $W=2$ strip mode. We note that for both pairs, the transfer matrix [Eq.~\eqref{eq: T W2}] is diagonal. Thus, the constraint from a (d, u)-pair anywhere in the strip, $v_{i, 2} = 0$, to satisfy the diagonal compatibility constraints [Eq.~\eqref{eq: plaquette constraint}] and lower strip condition [Eq.~\eqref{eq: BC bottom}] locally maps to the constraint $v_{1, 2} = 0$ on $\mathbf{v}_1$. Both (h, h)-pairs and (d, u)-pairs satisfy the diagonal compatibility constraints and lower strip condition locally with this constraint, see Fig.~\ref{fig: width 2}(f) for an example strip deformation. To result in valid strip mode, we also require the periodic strip condition [Eq.~\eqref{eq: PBC}] to be satisfied. We find that $v_{1, 2} = v_{k+1, 2} = 0$ and \ch{$v_{1, 1} = (-1)^{\mathrm{No.} \mathrm{(h, h)}} v_{k+1, 1}$, where $\mathrm{No.} \mathrm{(h, h)}$} is the number of (h, h)-pairs in the strip with periodic boundary conditions, thereby satisfying the periodic strip condition [Eq.~\eqref{eq: PBC}]. 

Thus, a strip that consists of any number of (h, h)-pairs and at least one (d, u)-pair 
satisfies all constraints as well as the CC and NT conditions when $\mathbf{v}_1 = (v_{1, 1}, 0)$ with $v_{1, 1} \neq 0$, thereby resulting in a valid $W=2$ strip mode.
Hence, the pairing rules for configurations that support valid $W=2$ strip modes are the following:
\begin{enumerate}[(i)]
    \item Every $2\times 2$ configuration of building blocks in the strip must be an (h, h)-pair or (d, u)-pair.
    \item There must be at least a single (d, u)-pair in the strip.
\end{enumerate}
These are sufficient conditions to obtain $W=2$ strip modes. They can also be shown to be necessary conditions, because any pair that is not a (h, h)-pair or (d, u)-pair constrains the strip deformation $\mathbf{v}_1$ to $v_{1, 1} = 0$, or $v_{1, 1} = -v_{1, 2}$, or both (see Appendix~\ref{App: Diag W2}), thereby breaking the nontrivial (NT) condition and therefore does not result in a valid $W=2$ strip mode [Fig.~\ref{fig: width 2}(g)]. Hence, these pairing rules are necessary and sufficient conditions on the strip configuration to obtain $W=2$ strip modes.

\subsection{Case 3: $W=3$}
Finally, we consider strips of width $W=3$. We show that in addition to simple adjustments to the pairing rules, we require an additional rule restricting the ordering of pairs in the strip configuration. This ordering rule highlights that the problem of constructing configurations that support valid strip modes is not reducible to a tiling problem which relies on nearest-neighbor interactions, but rather requires information of the entire strip configuration. This is surprising, as these rules emerge from local compatibility constraints. The new set of rules that we obtain are necessary and sufficient conditions to obtain $W=3$ strip modes.

First, we extend our transfer mapping to account for the extra row of building blocks in the strip. As in the previous two cases, we set the amplitude $\alpha^u=0$ such that the deformations of column $i$ are completely determined by fixing vector $\mathbf{v}_i = (v_{i, 1}, v_{i, 2}, v_{i, 3})$. We again want to obtain a complete map from $\mathbf{v}_i$ to $\mathbf{v}_{i+1}$. The maps for $v_{i+1, 1}$ and $v_{i+1, 2}$ do not depend on the extra row of building blocks and therefore follow Eq.~\eqref{eq: W1 map} and Eq.~\eqref{eq: W2 map} respectively. To obtain a map for $v_{i, 3}$, we implement the horizontal and vertical compatibility constraints [Eq.~\eqref{eq: compatibility constraints}] and find a linear mapping for $v_{i+1, 3}$ (see Appendix~\ref{App: transfer mapping}):
\begin{eqnarray}
    v_{i+1, 3} &= \frac{L_u(c_{i+1, 2})}{L_v(c_{i+1, 2})} \frac{L_u(c_{i+1, 3})}{L_v(c_{i+1, 3})} \left ( \frac{R_u(c_{i, 2})}{L_u(c_{i+1, 2})} - \frac{R_v(c_{i, 1})}{L_v(c_{i+1, 1})} \right) v_{i, 1} \nonumber\\
    & + \frac{L_u(c_{i+1, 3})}{L_v(c_{i+1, 3})} \left ( \frac{R_u(c_{i, 3})}{L_u(c_{i+1, 3})} - \frac{R_v(c_{i, 2})}{L_v(c_{i+1, 2})}  \right ) v_{i, 2} \nonumber \\
    & - \frac{R_v(c_{i, 3})}{L_v(c_{i+1, 3})} v_{i, 3}~.
    \label{eq: W3 map}
\end{eqnarray}
Together, Eq.~\eqref{eq: W1 map}, Eq.~\eqref{eq: W2 map} and Eq.~\eqref{eq: W3 map} form the transfer mapping from $\mathbf{v}_i$ to $\mathbf{v}_{i+1}$, which we capture compactly as $\mathbf{v}_{i+1} = T(\mathbf{c}_i, \mathbf{c}_{i+1}) \mathbf{v}_{i}$. Note that the transfer matrix $T(\mathbf{c}_{i}, \mathbf{c}_{i+1})$ is now a $3\times 3$ lower-triangular matrix that depends on the orientations $\mathbf{c}_i = (c_{i, 1}, c_{i, 2}, c_{i, 3})$ of the building blocks in column $i$ and column $i+1$.

Now, we want to find values for the orientations $\mathbf{c}_i$ that produce a valid strip mode, i.e., a deformation $\mathbf{v}_1$ which satisfies all constraints: the diagonal compatibility constraints [Eq.~\eqref{eq: plaquette constraint}], lower strip condition [Eq.~\eqref{eq: BC bottom}] and periodic strip condition [Eq.~\eqref{eq: PBC}]. Additionally, the strip deformation $\mathbf{v}_1$ should satisfy the CC and NT conditions. We note that $\mathbf{v}_1= 0$ corresponds to the strip deforming as an area of CRS with zero amplitude, i.e., not deforming at all. Additionally, $v_{1, 1} = 0$ with $v_{1, 2} \neq 0$ and $v_{1, 3} \neq 0$ corresponds to the top row 
not deforming at all
and $v_{1, 2} = -v_{1, 3}$ with $v_{1, 1} \neq 0$ corresponds to the bottom row deforming as an area of CRS with arbitrary amplitude. All these cases break the nontrivial (NT) condition as they describe strip deformations completely or in-part composed of rows of CRS blocks and thus do not describe valid $W=3$ strip modes. We exclude these configurations.

\begin{figure*}[t]
    \centering
    \includegraphics{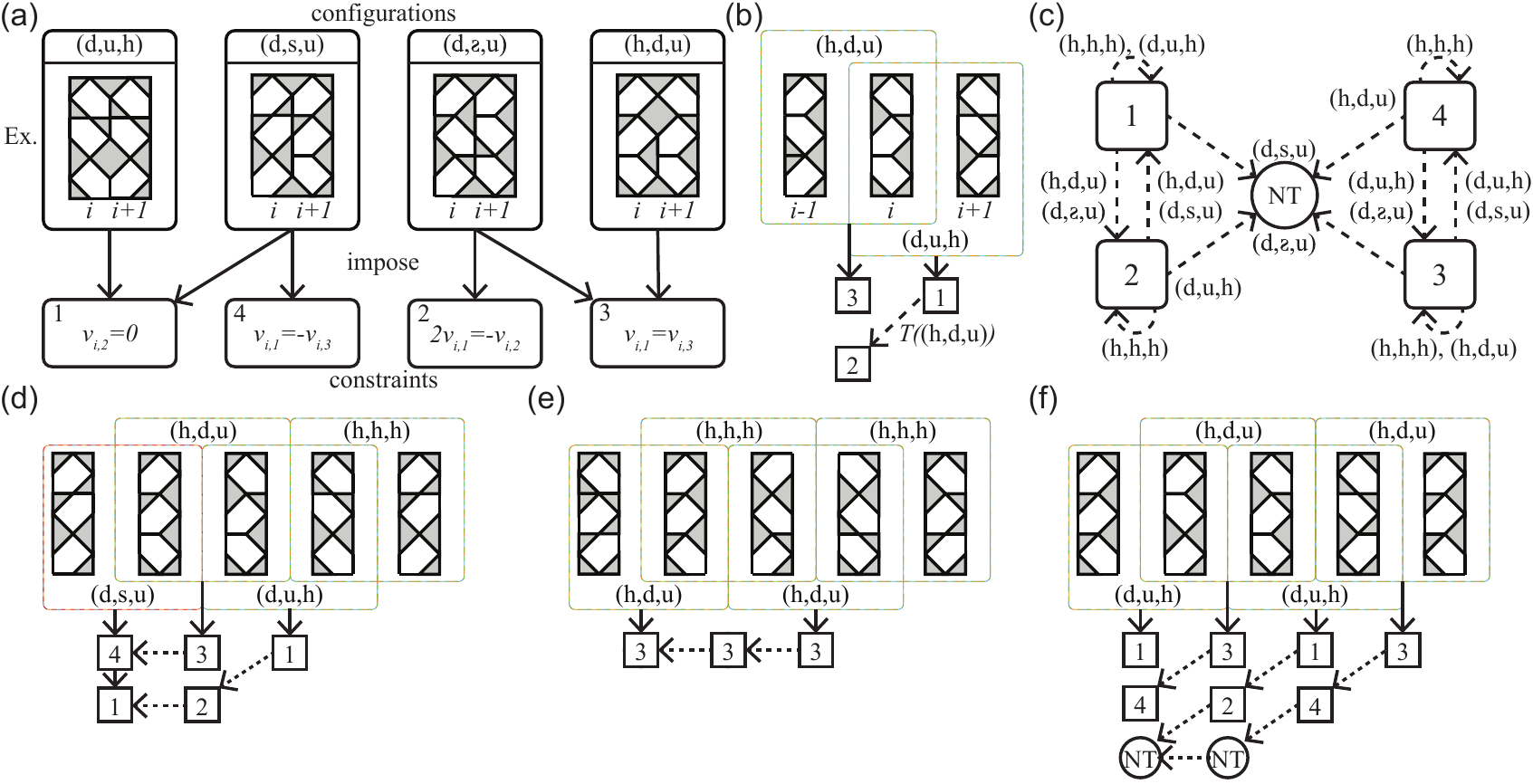}
    \caption{
    (a) $2\times3$ configurations $(\mathbf{c}_i, \mathbf{c}_{i+1})$ consisting of triplets of (d, u, h)-pairs, (d, s, u)-pairs, (d, \reverses, u)-pairs and (h, d, u)-pairs, impose one or two of four constraints (Eqs.~\eqref{eq: w3 const 1}-\eqref{eq: w3 const 4}, labeled 1 to 4 respectively) on the local strip deformation $\mathbf{v}_i$. We indicate this by black solid arrows. Note that the shown configurations are examples of the indicated pair type; other configurations that belong to the same type are possible.
    (b) A tiling of a (h, d, u)-pair and (d, u, h)-pair. Both pairs impose a constraint on their local strip deformation, $\mathbf{v}_{i-1}$ and $\mathbf{v}_i$ respectively, indicated by solid black arrows pointing to squares with numbers corresponding to the constraints as indicated in (a). Additionally, the constraint on $\mathbf{v}_i$ can be mapped using the transfer matrix $T(\mathrm{(h,d,u)})$ to a constraint on $\mathbf{v}_{i-1}$. This imposes constraint 2 with $i\mapsto i-1$ on $\mathbf{v}_{i-1}$, the transfer mapping is indicated by the dashed arrow.
    (c) The constraints map from a constraint on strip deformation $\mathbf{v}_i$ to a constraint on $\mathbf{v}_{i-1}$ under application of the transfer mapping $T(\mathbf{c}_{i-1}, \mathbf{c}_{i})$ (dashed arrows) for the configurations $(\mathbf{c}_{i-1}, \mathbf{c}_{i})$ as indicated next to the arrows. Note that here we only consider (h, h, h)-pairs, (d, u, h)-pairs, (h, d, u)-pairs, (d, s, u)-pairs and (d, \reverses, u)-pairs. A constraint maps to the indicated constraint with $i\mapsto i-1$. Constraints are labeled by number as indicated in (a). Every constraint can map to a constraint that breaks the NT condition. 
    (d) Example of a valid $k=4$ strip configuration supporting a $W=3$ strip mode. Notice that we take periodic boundary conditions. There are two mapped and local constraints on $\mathbf{v}_1$, thereby satisfying the CC condition, and both results do not break the NT condition, resulting in a valid $W=3$ strip mode.
    (e) Example of an invalid strip configuration that does not support a valid $W=3$ strip mode. There is only one mapped and local constraint on $\mathbf{v}_1$, breaking the CC condition. The strip deformation can be decomposed into a $W=1$ strip mode and $W=2$ strip mode.
    (f) Example of an invalid strip configuration that does not support a valid $W=3$ strip mode. Some constraints map to a constraint that breaks the NT condition, resulting in an invalid strip deformation.}
    \label{fig:w3 constraint mappings}
\end{figure*}

To construct valid strip configurations, we consider $2\times 3$ configurations of building blocks $(\mathbf{c}_i, \mathbf{c}_{i+1})$. Again, we compose such configurations by vertically stacking pairs of horizontally adjacent building blocks $(c_{i, j}, c_{i+1, j})$ for the top row $j=1$, middle row $j=2$ and bottom row $j=3$, e.g., a triplet of d-, u-, and h-pairs, which we denote as a (d, u, h)-pair, satisfies condition [Eq.~\eqref{eq: cond d}] for $(c_{i, 1}, c_{i+1, 1})$, satisfies condition [Eq.~\eqref{eq: cond u}] for $(c_{i, 2}, c_{i+1, 2})$ and satisfies condition [Eq.~\eqref{eq: cond h}] for $(c_{i, 3}, c_{i+1, 3})$, see Fig.~\ref{fig:w3 constraint mappings}(a) for an example of a (d, u, h)-pair. 
Additionally, we now distinguish between the s-pair $(c_{i, j}, c_{i+1, j}) = (\mathrm{NE}, \mathrm{SW})$ and the \reverses-pair $(c_{i, j}, c_{i+1, j}) = (\mathrm{SE}, \mathrm{NW})$ [Fig.~\ref{fig: width 2}(b)] despite both pairs satisfying condition [Eq.~\eqref{eq: cond s}] as configurations composed of such pairs impose distinct constraints on the local strip deformation $\mathbf{v}_i$.

In what follows, we will show that a valid strip configuration consists only of (h, h, h), (d, u, h), (h, d, u), (d, s, u) and (d, \reverses, u)-pairs. Specifically, we will show the following for strip configurations composed of such pairs:
\begin{enumerate}[(1)]
    \item each of these configurations except (h, h, h)-pairs imposes one or two constraints out of a set of four possible constraints on the deformation $\mathbf{v}_i$ to satisfy the diagonal compatibility constraints [Eq.~\eqref{eq: plaquette constraint}] and lower strip condition [Eq.~\eqref{eq: BC bottom}] locally.
    \item upon applying the transfer mapping $T(\mathbf{c}_{i-1}, \mathbf{c}_i)$, each of the four possible constraints on $\mathbf{v}_i$ map to constraints on $\mathbf{v}_{i-1}$ that are degenerate to the four possible constraints that can be imposed by the $(\mathbf{c}_{i-1}, \mathbf{c}_i)$-pair locally on $\mathbf{v}_{i-1}$ for most valid $2\times 3$ configurations. For the other valid configurations, the mapped constraints and local constraints imposed by the $(\mathbf{c}_{i-1}, \mathbf{c}_{i})$-pair on $\mathbf{v}_{i-1}$ together break the CC or NT conditions and do not result in a valid $W=3$ strip mode. We exclude such combinations. 
    \item constraints on the configurational ordering of $(\mathbf{c}_i, \mathbf{c}_{i+1})$-pairs are captured with a simple additional rule.
\end{enumerate}
We now consider these points one-by-one.

First, we find that for (d,u,h)-pairs, (h, d, u)-pairs, (d, s, u)-pairs, and (d, \reverses, u)-pairs the diagonal compatibility constraints [Eq.~\eqref{eq: plaquette constraint}] and lower strip condition [Eq.~\eqref{eq: BC bottom}] are satisfied locally by satisfying one or two of four different constraints on $\mathbf{v}_i$. These four different constraints are (see Appendix~\ref{App: Diag W3}): 
\begin{eqnarray}
    v_{i, 2} = 0~, \label{eq: w3 const 1}\\
    2 v_{i, 1} = -v_{i, 2}~, \label{eq: w3 const 2} \\
    v_{i, 1} = v_{i, 3}~, \quad \mathrm{and} \label{eq: w3 const 3} \\
    v_{i, 1} = -v_{i, 3}~. \label{eq: w3 const 4}
\end{eqnarray}
We find that a (d, u, h)-pair imposes constraint [Eq.~\eqref{eq: w3 const 1}], a (h, d, u)-pair imposes constraint [Eq.~\eqref{eq: w3 const 3}], a (d, s, u)-pair imposes constraints [Eq.~\eqref{eq: w3 const 1}] and [Eq.~\eqref{eq: w3 const 4}], and a (d, \reverses, u)-pair imposes constraints [Eq.~\eqref{eq: w3 const 2}] and [Eq.~\eqref{eq: w3 const 3}] on $\mathbf{v}_i$ [Fig.~\ref{fig:w3 constraint mappings}(a)]. An (h, h, h)-pair trivially satisfies the diagonal compatibility constraints and lower strip condition and does not place any constraints on $\mathbf{v}_i$.

Now, we combine the valid $2\times 3$ configurations (h, h, h)-pairs, (h, d, u)-pairs, (d, u, h)-pairs, (d, s, u)-pairs and (d, \reverses, u)-pairs in a strip configuration, see Fig.~\ref{fig:w3 constraint mappings}(b) for an example. We find that most combinations of these configurations result in a valid strip mode, but there are exceptions for which we devise a rule.
First, we consider each of the four constraints [Eqs. \eqref{eq: w3 const 1}-\eqref{eq: w3 const 4}] on $\mathbf{v}_{i}$ 
and use the transfer mapping $T(\mathbf{c}_{i-1}, \mathbf{c}_{i})$ to transform each constraint to a constraint on $\mathbf{v}_{i-1}$ for each valid $2\times 3$ configuration $(\mathbf{c}_i, \mathbf{c}_{i+1})$ (see Appendix~\ref{App: constraint mapping}).
The total set of constraints on $\mathbf{v}_{i-1}$ then consists of the mapped constraint(s) and local constraints imposed by the configuration $(\mathbf{c}_{i-1}, \mathbf{c}_{i})$ [Fig.~\ref{fig:w3 constraint mappings}(b)]. To have a valid strip mode, the total number of constraints must equal two to satisfy the CC condition. Additionally, none of the constraints may result in a strip deformation that does not satisfy the NT condition.

We find that the four constraints on $\mathbf{v}_{i}$ [Eqs.~\eqref{eq: w3 const 1}-\eqref{eq: w3 const 4}]
map within the set of these same four constraints with index $i \mapsto i-1$ on $\mathbf{v}_{i-1}$ for most configurations $(\mathbf{c}_{i-1}, \mathbf{c}_{i})$ (see Appendix~\ref{App: constraint mapping W3}, Fig.~\ref{fig:w3 constraint mappings}(c)).
However, for some configurations, the mapped constraints, when taken together with the local constraints imposed by the configuration on $\mathbf{v}_i$, result in a strip deformation $\mathbf{v}_i$ that breaks the NT condition [Fig.~\ref{fig:w3 constraint mappings}(f)]. 
To construct strip configurations that result in a valid $W=3$ strip mode we exclude combinations of valid configurations that result in such constraints.

We now aim to find what combinations of valid configurations do not result in a valid $W=3$ strip mode. The constraint mapping [Fig.~\ref{fig:w3 constraint mappings}(c)] prohibits certain combinations of valid configurations. In general, for a given strip configuration $C^{SM}$ each $(\mathbf{c}_i, \mathbf{c}_{i+1})$-pair imposes one or two constraints [Eqs.~\eqref{eq: w3 const 1}-\eqref{eq: w3 const 4}] on the local deformation $\mathbf{v}_i$. These constraints then need to be iteratively mapped to $\mathbf{v}_1$, starting from $\mathbf{v}_k$ [Fig.~\ref{fig:w3 constraint mappings}(d)-(f)]. If at any point in the strip configuration the CC or NT conditions on $\mathbf{v}_i$ are not satisfied, the strip configuration does not support a valid $W=3$ strip mode [Fig.~\ref{fig:w3 constraint mappings}(f)]. 
To find which sets of pairs result in invalid strip modes, we look for combinations of pairs that lead to a constraint on $\mathbf{v}_{i+1}$ that will get mapped to a constraint that breaks the NT condition on $\mathbf{v}_i$ using the constraint map [Fig.~\ref{fig:w3 constraint mappings}(c)]. 
We find that there are sets of pairs in either the top two rows or bottom two rows of the strip that are not allowed to occur in order anywhere in the strip (see Appendix~\ref{App: invalid sequences W3}). 
Moreover, this set of pairs can be freely padded with (h, h, h)-pairs as such pairs do not add any constraints of their own and act as an identity mapping for the constraints [Fig.~\ref{fig:w3 constraint mappings}(c)]. Thus, to determine if a strip configuration supports a valid strip mode requires knowledge of the entire strip configuration. 

\begin{figure}[t]
    \centering
    \includegraphics{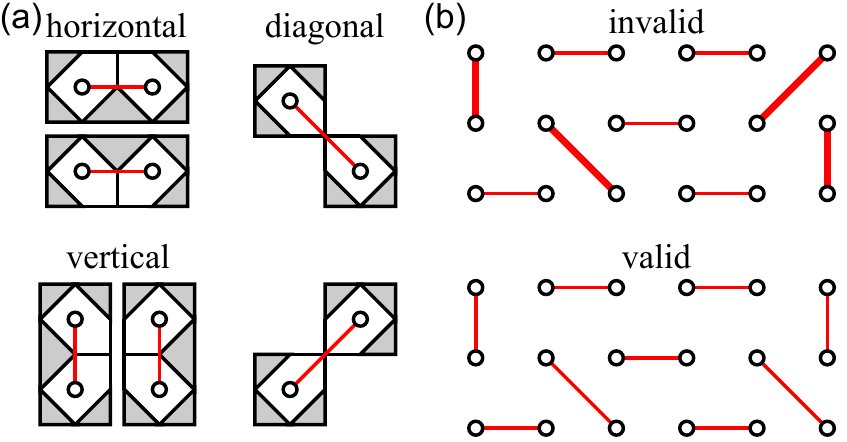}
    \caption{(a) Linked pairs of building blocks are marked by a \ch{white} circle in the center of each building block and a red \ch{solid} line that connects the circles. Linked building blocks are labeled by orientation.
    (b) $k=6$ strip configurations are represented as collections of linked building blocks. The invalid configuration (top) breaks rule \eqref{rule: general 2} as it contains both vertically and diagonally linked building blocks in both pairs of adjacent rows (thick red \ch{solid} lines). The valid configuration (bottom) describes a strip configuration that supports a $W=3$ strip mode.}
    \label{fig:linked rules}
\end{figure}

We observe that the combinations of valid configurations that result in an invalid strip mode all follow a simple configurational rule. To formulate this rule, we note that the nontrivial diagonal edge $d^c$ of each building block in a strip composed of valid configurations meets at a vertex with a single other nontrivial diagonal edge of a building block in the strip. We refer to such pairs of building blocks as \textit{linked}. Linked building blocks can be oriented either horizontally, vertically or diagonally with respect to each other [Fig.~\ref{fig:linked rules}(a)]. 
We observe that sequences of valid configurations that result in an invalid strip mode always contain both vertically linked and diagonally linked building blocks.  
Thus we can formulate a simple rule to exclude invalid sequences: all building blocks linked together in two adjacent rows can only be linked vertically or diagonally, never both.

We capture these necessary requirements in a compact set of design rules:
\begin{enumerate}[(i)]
    \item Every $2\times 3$ configuration of building blocks in the strip must be a (h, h, h)-pair, (d, u, h)-pair, (h, d, u)-pair, (d, s, u)-pair or (d, \reverses, u)-pair. \label{rule: w3 1}
    \item There must be at least a single d-pair in the top row and at least a single u-pair in the bottom row. \label{rule: w3 2}
    \item All linked building blocks in two adjacent rows can only be linked vertically and horizontally or diagonally and horizontally. \label{rule: w3 3}
\end{enumerate}
Rule~\eqref{rule: w3 2} is required to satisfy the constraint counting (CC) condition and result in a single $W=3$ strip mode, rather than multiple smaller strip modes [Fig.~\ref{fig:w3 constraint mappings}(e)]. Rule~\eqref{rule: w3 3} is added to exclude invalid sequences of configurations that do not result in a valid $W=3$ strip mode [Fig.~\ref{fig:w3 constraint mappings}(f)]. Note that this rule is global---checking it requires knowledge of the entire strip.
This is because the CC condition now permits two constraints, both of which can potentially map to a constraint that breaks the nontrivial (NT) condition. A constraint introduced at the very end of the strip can be mapped throughout the entire strip and only encounter an incompatible configuration at the beginning of the strip.
These are sufficient conditions to obtain $W=3$ strip modes. They can also be shown to be necessary conditions, because other $2\times 3$ configurations constrain the strip deformation to $v_{1, 1} = 0$, $v_{1, 1} = -v_{1, 2}$ or $v_{1, 2} = -v_{i, 3}$ or combinations, thereby breaking the NT condition and therefore do not result in a valid $W=3$ strip mode (see Appendix~\ref{App: Diag W3}). Hence, these pairing rules are necessary and sufficient conditions on the strip configuration to support a $W=3$ strip mode.

\subsection{Towards general design rules\label{sec:general design rules}}
\begin{figure*}[t]
    \centering
    \includegraphics{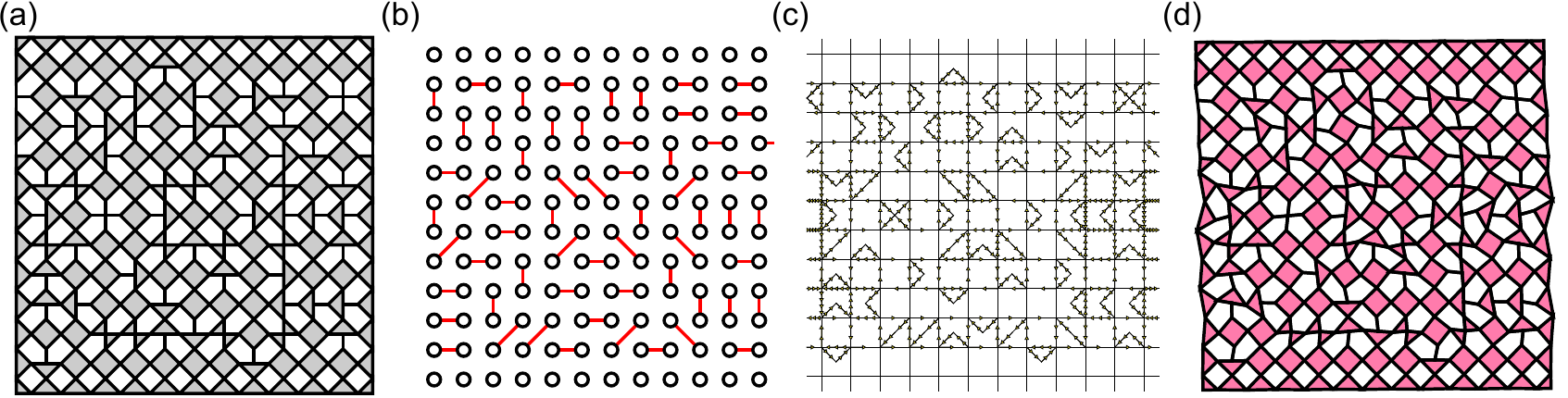}
    \caption{Realization of a $12 \times 12$ unit cell that supports a $W=10$ strip mode. \ch{To better illustrate the kinematics of the strip mode, we have restrained the top and bottom layers of building blocks to deform solely with the CRS mode.} (a) Schematic representation of the unit cell. (b) Linked pairs of the configuration. Note that the horizontal strip between rows 2 and 11 satisfies the general design rules. (c) Vertex representation of the strip mode. The number of arrows on horizontal and vertical edges connecting two building blocks is reduced by half for clearer visualization. (d) Schematic representation of the unit cell deforming as the strip mode.}
    \label{fig:strip W10}
\end{figure*}

Now we discuss how these design rules generalize to larger width $W$ strip configurations. We have proven that the rules we found for strip modes of width $W=1$, $W=2$ and $W=3$ are necessary and sufficient requirements on a strip configuration to support a valid strip mode. 
Based on these rules, we formulate a general set of rules that we conjecture are, at the least, also sufficient requirements for larger width $W$ strip modes. 
We formulate these rules completely in terms of linked building blocks [Fig.~\ref{fig:linked rules}(a)]:
\begin{enumerate}[(i)]
    \item Every building block in the strip must be linked with a single other building block in the strip \label{rule: general 1}
    \item All linked building blocks in two adjacent rows must only be linked vertically and horizontally or diagonally and horizontally, never vertically and diagonally. \label{rule: general 2}
\end{enumerate}
The smallest width $W$ and irreducible strip in the unit cell for which these rules hold supports a strip mode of width $W$. Rule~\eqref{rule: general 2} is a global rule; checking it requires knowledge of the entire strip [Fig.~\ref{fig:linked rules}(b)]. We find a perfect agreement of our rules for $\sim 10^6$ randomly generated $k\times k$ unit cell designs to be of class (iii) or not (see Appendix~\ref{App: numerical proof rules}).
We therefore have strong numerical evidence that our rules are not only necessary to have a strip mode, but also that strip modes are the only type of zero mode that result in class (iii) mode-scaling.
As a final indication that these rules are sufficient for a strip configuration to support a strip mode, we use the rules to design a strip mode of width $W=10$ [Fig.~\ref{fig:strip W10}]. 

\section{Discussion}
The rational design of multiple soft modes in aperiodic metamaterials is intrinsically different from tiling- or spin-ice based design
strategies for a single soft mode~\cite{coulais2016combinatorial, nevzerka2018jigsaw, dieleman2020jigsaw, meeussen2020topological, bossart2021oligomodal, mastrigt2022machine}. The key challenge is 
to precisely control the
balance between the kinematic degrees of freedom with the kinematic constraints.
For increasing sizes, these constraints proliferate though the sample, and to obtain multiple soft modes, the spatial design must be such that many of these constraints are degenerate. What is particularly vexing is that these constraints act on a growing set of 
local kinematic degrees of freedom, so that  checking for degenerate constraints is cumbersome. As a consequence, current design strategies for multimodal metamaterials rely on computational methods, in either continuous systems~\cite{kim2019conformational, choi2019programming, choi2021compact} or discrete systems~\cite{dykstra2023inverse}.

Here, we introduced a general transfer matrix-like framework for mapping the local constraints to a small, pre-defined subset of kinematic degrees of freedom, and 
use this framework to obtain effective tiling rules for a combinatorial multimodal metamaterial. Strikingly, beside the usual local rules which express constraints on pairs of adjacent building blocks, we find nonlocal rules that restrict the types of tiles that are allowed to appear together {\em anywhere} in the metamaterial. These kind of nonlocal rules are unique to multimodal metamaterials.

More broadly, our work is a first example
where metamaterial design leads to 
complex combinatorial tiling problems that are beyond the limitations of Wang tilings.
It is complementary to combinatorial computational methods used in design of irregular architectured materials~\cite{liu2022growth} or computer graphics~\cite{bian2018tile} that use local tiling rules to fabricate complicated spatial patterns. 

Conversely, instead of clear-cut local rules that state which tiles fit together, our method requires careful bookkeeping of local constraints imposed by placed tiles and propagation of these constraints through all previously placed tiles to a single set of degrees of freedom. As a result, knowledge of a tile's neighbors is no longer sufficient information to determine if that tile can be placed. Instead, one requires knowledge of most, if not all, previously placed tiles. 
We believe our method is well-suited to tackle tiling problems beyond Wang tiles. Several open questions remain: are nonlocal rule generically emerging in multimodal metamaterials? How does our method relate to other emergent nonlocal tiling constraints that arise, for example, in the fields of computer graphics~\cite{merrell2007example, merrell2008continuous, merrell2010model} and chip design~\cite{adya2005combinatorial, oh2022bayesian}?.
Additionally, our method is limited to design of zero modes and thus may be insufficient when designing for larger deformations. How to adjust our method to include nonlinear kinematic constraints is an open question.

Our framework opens up a new route for rational design of spatially textured soft modes in multimodal metamaterials, which we demonstrate 
by designing metamaterials with strip modes of targeted width and location. Such strip modes can be utilized to control buckling and energy-absorption under uniaxial compression perpendicular to the orientation of the strip~\cite{liu2023leveraging}. Our method can readily be extended to edge-modes, by considering, e.g., horizontal edge strips, imposing the upper strip condition and periodic strip condition and taking into account open boundary conditions at the bottom of the strip. Similarly, Swiss cheese-modes can be modeled by imposing upper and lower strip conditions horizontally and vertically at appropriate locations in the metamaterial. 
Additionally, our method can be extended to design three dimensional metamaterials by constructing an additional transfer matrix that propagates local degrees of freedom (dof) along the newly added spatial dimension. To ensure kinematic compatibility, additional constraints may need to be introduced to ensure different dof propagation paths result in the same final deformation.
We hope our work will push the interest in multimodal metamaterials whose mechanical functionality is selectable through actuation, with potential applications in programmable materials, soft robotics, and computing {\em in materia}.

\begin{acknowledgments}
\textit{Data availability statement.}\textemdash The code supporting the findings reported in this paper is publicly available on GitLab~\footnote{See \url{https://uva-hva.gitlab.host/published-projects/CombiMetaMaterial} for code to calculate zero modes and numerically check design rules. \label{fn: code}}---the data on Zenodo~\cite{Zenodo_MetaCombi}.

\textit{Acknowledgments.}\textemdash We thank David Dykstra and Marjolein Dijkstra for discussions. C.C. acknowledges funding from the European Research Council under Grant Agreement No. 852587. 
\end{acknowledgments}

\bibliography{bibliography}
\appendix
\section{Open boundary conditions \label{App: open boundary conditions}}
Here, we show that angles located at an open boundary can deform unconstrained, both at the faces of the building blocks ($u, v, l, r$) and corners ($d^c$). First, we consider angles at the face of each building block. If the face of the building block is located at an open boundary, the angle can deform freely as there is no competing adjacent angle. For example, if the top face of a building block $z$ is located at an open boundary, there are no constraints placed upon deformation $u_z$.

Second, we consider the nontrivial corner angle $d^c$ of a building block with orientation $c$ where the corner is located at an open boundary. Here, there can be an adjacent diagonal angle of a neighboring building block. However, the diagonal angle is still unconstrained in its deformation at the open boundary, regardless if it is adjacent to a diagonal angle of another building block. To see this, note that for the two neighboring building blocks at an open boundary to be kinematically compatible, only the angles at the shared face between the two building blocks are constrained with the horizontal or vertical compatibility constraint. For example, two horizontally neighboring building blocks at locations $z$ and $z+1$ with their top face at the open boundary deform compatibly only if the right and left angles satisfy $r_z = -l_{z+1}$. More formally, this can be shown by composing the compatibility matrix for these two building blocks in all possible orientations and determining the dimension of the matrix' null space~\cite{calladine1978buckminster, lubensky2015phonons}. This dimension is always equal to six, which corresponds to three floppy modes and three trivial modes: rotation and translation. As each building block has two zero modes, there must only be one constraint placed on their deformations: the horizontal compatibility constraint. As there are no states of self-stress in this structure, the number of floppy modes also follows from a simple Maxwell counting argument~\cite{maxwell1864calculation}. Thus, nontrivial diagonal corners $d^c$ located at the open boundary can deform unconstrained.

\section{Realizations mode structure \label{App: modeshape realizations}}
\begin{figure}[b]
    \centering
    \includegraphics{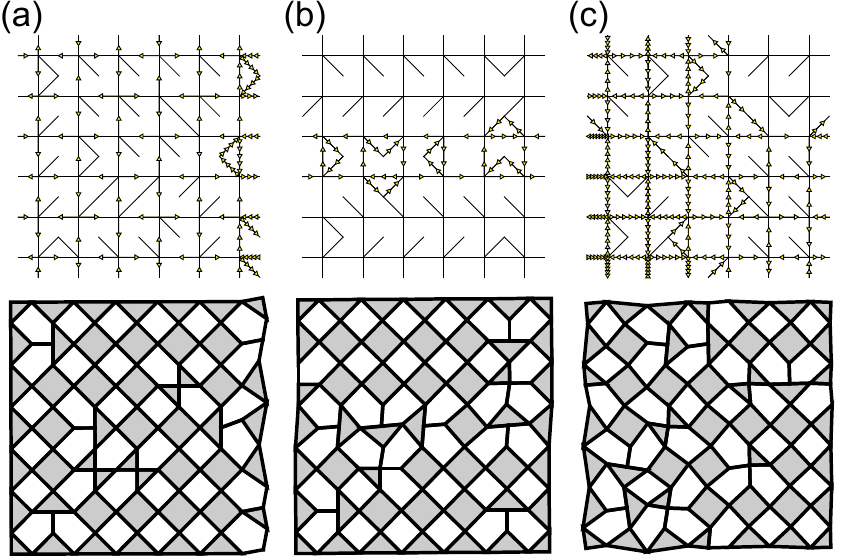}
    \caption{Vertex (top) and schematic (bottom) representations of an edge-mode (a), strip mode (b) and Swiss cheese-mode (c). \ch{Note that we replaced rigid pentagons with a reentrant edge with rigid diamonds (rotated squares) that are kinematically equivalent in the schematic representation for ease of interpretation.}}
    \label{fig:modeshape realizations}
\end{figure}

Here we show explicit realizations of unit cells that support an edge-mode [Fig.~\ref{fig:modeshape realizations}(a)], a strip mode [Fig.~\ref{fig:modeshape realizations}(b)] and a Swiss cheese-mode [Fig.~\ref{fig:modeshape realizations}(c)] as described in Sec.~\ref{sec: mode structure} and Fig.~\ref{fig:modeshapes}.

\section{Linear coordinate transformations \label{App: coordinate transform}}
To find conditions on the strip configuration $C^{SM}$, we change to a more convenient basis where instead of mode amplitudes $\alpha_z$ and $\beta_z$, deformations $u_z$ and $v_z$ are the two degrees of freedom for each building block. We find
\begin{equation}
    \begin{pmatrix} u_z \\ v_z \end{pmatrix} = \begin{bmatrix}
        1   &&  u_{D}(c_z)\\
        1   &&  v_{D}(c_z)
    \end{bmatrix} \begin{pmatrix}
        \alpha_z \\
        \beta_z
    \end{pmatrix} = \Lambda(c_z) \begin{pmatrix} \alpha_z \\ \beta_z \end{pmatrix},
\end{equation}
where $u_{D}(c_z)$ and $v_{D}(c_z)$ are the $u$- and $v$-components of the basis zero mode $m_{D}(c_z)$.
Subsequently, we invert $\Lambda(c_z)$ to find the change of basis matrix
\begin{equation}
    \Lambda^{-1}(c_z) = \frac{1}{v_{D}(c_z) - u_{D}(c_z)} \begin{bmatrix}
        v_{D}(c_z)  &&  -u_{D}(c_z) \\
        -1      &&  1
    \end{bmatrix},
\end{equation}
which is well-defined since $u_{D}(c_z)\neq v_{D}(c_z)$ for all orientations $c_z$.
We can express deformations $l_z$, $r_z$, and $d_z^o$ in terms of $(u_z, v_z)$ using the change of basis matrix:
\begin{eqnarray}
    \begin{pmatrix}
    l_z \\ r_z
    \end{pmatrix} = \Gamma(c_z) \Lambda^{-1}(c_z) \begin{pmatrix} u_z \\ v_z \end{pmatrix}
\end{eqnarray}
and
\begin{eqnarray}
    \quad d_z^{o} = \zeta^{o}(c_z) \Lambda^{-1}(c_z) \cdot \begin{pmatrix} u_z \\ v_z \end{pmatrix},
\end{eqnarray}
where
\begin{align}
    \Gamma(c_z) &= \begin{pmatrix}
        -1 & l_{D}(c_z) \\
        -1 & r_{D}(c_z)
    \end{pmatrix} \quad \mathrm{and} \\
    \zeta^{o}(c_z) &= (0, d_{D}^{o}(c_z))
\end{align}
depend on the orientation $c_z$ of the building block. Note that $o$ denotes the orientation of diagonal deformation $d_z^{o}$, which is independent from the building block orientation $c_z$. The equation for $l_z, r_z$ and $d_z^o$ further simplify to
\begin{eqnarray}
    l_z = L_u(c_z)\,u_z + L_v(c_z)\,v_z, \\
    r_z = R_u(c_z)\,u_z + R_v(c_z)\,v_z
\end{eqnarray}
and
\begin{eqnarray}
    d^o_z = D^o(c_z)(-u_z + v_z).
\end{eqnarray}
Values of the coefficients $L_u, L_v, R_u, R_v, D^o$ for the four orientations $c_z$ are given in Table~\ref{tab:L, R, D}.

\section{Deriving the transfer mapping \label{App: transfer mapping}}
Here we derive the linear transfer mapping that maps the vertical deformations $\mathbf{v}_{i} = (v_{i, 1}, v_{i, 2}, ..., v_{i, W})$ of row $i$ in a strip configuration of width $W$ to the vertical deformations $\mathbf{v}_{i+1}$ of the neighboring row $i+1$ such that $\mathbf{v}_{i+1} = T(\mathbf{c}_i, \mathbf{c}_{i+1}) \mathbf{v}_{i}$. We consider a $2\times W$ strip configuration of unspecified orientations $(\mathbf{c}_i, \mathbf{c}_{i+1})$. To derive this transfer matrix, we solve the horizontal and vertical compatibility constraints [Eq.~\eqref{eq: compatibility constraints}] and upper boundary condition [Eq.~\eqref{eq: BC top}] iteratively for the vertical deformations $\mathbf{v}_i$ [Fig.~\ref{fig:T schematic}(a)-(f)]. We first consider the first row in the strip such that $j=1$. From the upper boundary conditions and setting $u_{i, 1} = 0$ without loss of generality, we find $u_{i, 1} = - u_{i+1, 1} = 0$ such that the horizontal compatibility constraint reduces to
\begin{equation}
    v_{i+1, 1} = - \frac{R_v(c_{i, 1})}{L_v(c_{i+1, 1})} v_{i, 1}.
    \label{eq: v_{i+1, 1} to v_{i, 1}}
\end{equation}
We now consider a general row $j$ and find that the horizontal compatibility condition reduces to
\begin{eqnarray}
    v_{i+1, j} = \frac{R_u(c_{i, j})}{L_v(c_{i+1, j})} v_{i, j-1} - \frac{R_v(c_{i, j})}{L_v(c_{i+1, j})} v_{i, j} \nonumber\\
    + \frac{L_u(c_{i+1, j})}{L_v(c_{i+1, j})} v_{i+1, j-1}.
    \label{eq: v_i+1, j to v_i+1, j-1}
\end{eqnarray}
Thus we can solve for $v_{i+1, j}$ in terms of $\mathbf{v}_i$ by recursively applying this equation.
We find
\begin{eqnarray}
    v_{i+1, j} = \sum_{a=1}^{j-1} \left(\frac{R_u(c_{i, a+1})}{L_u(c_{i+1, a+1})} - \frac{R_v(c_{i, a})}{L_v(c_{i+1, a})}\right) \nonumber\\ 
    \times \prod_{b=a+1}^{j}\frac{L_u(c_{i+1, b})}{L_v(c_{i+1, b})} v_{i, a} - \frac{R_v(c_{i, j})}{L_v(c_{i+1, j})} v_{i, j},\
    \label{eq: v_i+1 to v_i map}
\end{eqnarray}
for $j\geq 2$. The linear map from $\mathbf{v}_{i}$ to $\mathbf{v}_{i+1}$ is captured in the transfer matrix $T(\mathbf{c}_i, \mathbf{c}_{i+1})$, which we can now define explicitly:
\begin{widetext}
\begin{equation}
    T(\mathbf{c}_i, \mathbf{c}_{i+1})_{a, b} = \begin{cases}
        \left(\frac{R_u(c_{i, b+1})}{L_u(c_{i+1, b+1})} - \frac{R_v(c_{i, b})}{L_v(c_{i+1, b})}\right) \prod_{j=b+1}^{a}\frac{L_u(c_{i+1, j})}{L_v(c_{i+1, j})}, &\mathrm{if} \quad b<a \\
        - \frac{R_v(c_{i, b})}{L_v(c_{i+1, b})}, &\mathrm{if} \quad b = a \\
        0, &\mathrm{if} \quad b > a.
    \end{cases}
    \label{eq: T matrix general}
\end{equation}
\end{widetext}

\section{nontrivial conditions \label{App: NT conditions}}
Here we show that a CRS site, which has deformations $u_{i, j} = v_{i, j}$, in the top row of the strip SM or bottom row of the strip leads to that entire strip deforming with the CRS mode, breaking the nontrivial (NT) condition. This already follows from the restriction on the mode structure described in Sec.~\ref{sec: mode structure}, but for completeness we derive it here using the transfer-matrix formalism. Moreover, we show that a CRS block in the second row of the strip in a $W=3$ strip breaks the NT condition. We first consider a CRS site in the top row, such that $j=1$, and use the upper boundary conditions [Eq.~\eqref{eq: BC top}] and transfer matrix [Eq.~\eqref{eq: T matrix general}] to show that the left-most vertical deformation $v_{1, 1} = 0$. Second, we consider a CRS site in the bottom row, such that $j=W$, and use lower boundary conditions [Eq.~\eqref{eq: BC bottom}] and transfer matrix [Eq.~\eqref{eq: T matrix general}] to show that the left-most vertical deformations $v_{i, W-1} = -v_{i, W}$. Finally, we consider a CRS site in the second row, such that $j=2$, and use the transfer matrix [Eq.~\eqref{eq: T matrix general}] to show that such a block results in a constraint on $\mathbf{v}_{1}$ that is incompatible with the four $W=3$ constraints [Eqs.~\eqref{eq: w3 const 1}-\eqref{eq: w3 const 4}] and breaks the NT condition.

First, we consider a general strip configuration $C^{SM}$. Suppose the building block at site $(i, 1)$ can only deform with the CRS mode, such that deformations $u_{i, 1} = v_{i, 1}$. Without loss of generality, we set the left-most upper deformation $u_{1, 1} = 0$. From the upper boundary condition [Eq.~\eqref{eq: BC top}] we find that upper deformation $u_{i, 1} = 0$, such that the vertical deformation becomes $v_{i, 1} = 0$. The transfer matrix [Eq.~\eqref{eq: T matrix general}] is lower triangular, thus $v_{i, 1}$ only depends on the upper left-most vertical deformation $v_{1, 1}$ by a factor consisting of the product of the diagonal transfer matrix components $T(\mathbf{c}_i, \mathbf{c}_{i+1})_{1, 1}$ of the building block pairs between $(1, 1)$ and $(i, 1)$. Therefore, if $v_{i, j} = 0$, $v_{1, 1} = 0$ must be true as well. Moreover, $v_{a, 1} = 0$ for all columns $a$ in the strip. Thus all building blocks in the strip are CRS sites and deform with $u_{i, 1} = v_{i, 1} = 0$, resulting in the entire top row of the strip deforming as a CRS mode compatibly with area of CRS with amplitude $\alpha^u=0$.

Second, we consider a general strip configuration $C^{SM}$ where the building block at site $(i, W)$ deforms with the CRS mode, such that $u_{i, W} = v_{i, W}$. From the vertical compatibility constraint [Eq.~\eqref{eq: compatibility constraints}] we know that $u_{i, W} = -v_{i, W-1}$, such that $v_{i, W-1} = - v_{i, W}$. To find the deformations of the left neighbor at site $(i+1, W)$, we plug the map [Eq.~\eqref{eq: v_i+1 to v_i map}] into the lower boundary condition [Eq.~\eqref{eq: BC bottom}] to find
\begin{equation}
    v_{i+1, W} \frac{L_v(c_{i+1, W}) - R_u (c_{i, W}) - R_v(c_{i, W})}{L_u(c_{i+1, W})} = -v_{i+1, W-1}.
\end{equation}
The fraction reduces to 1 for all possible configuration pairs $(c_{i, W}, c_{i+1, W})$ (see Table~\ref{tab:L, R, D}), such that the vertical deformations of the neighboring building block deform as $v_{i+1, W} = -v_{i+1, W-1}$ and the block is a CRS site. Similarly, we can do the same calculation for the right neighbor at site $(i-1, W)$ to find
\begin{equation}
    v_{i-1, W-1} = v_{i-1, W} \frac{R_v (c_{i-1, W}) - L_u(c_{i, W}) - L_v(c_{i, W})}{R_u(c_{i-1, W})}
\end{equation}
which also reduces to $v_{i-1, W-1} = - v_{i-1, W}$ for all possible configuration pairs $(c_{i-1, W}, c_{i, W})$. Thus we find that a CRS site in the bottom of the strip results in CRS sites to its left and right neighbor upon requiring the lower boundary condition [Eq.~\eqref{eq: BC bottom}] to be satisfied. In conclusion, we find that a single CRS site in the top or bottom row of the strip SM results in that entire row deforming as an area of CRS, breaking the NT condition.

Next, we consider how a CRS site in row $j=2$, where $v_{i, 1} = -v_{i, 2}$, maps to a constraint on $\mathbf{v}_{i-1}$. This mapping depends on the configuration of columns $(\mathbf{c}_{i}, \mathbf{c}_{i+1})$. In general, we find it maps to
\begin{eqnarray}
    v_{i-1, 1} = -\frac{T_{2, 2}}{T_{1, 1} + T_{2, 1}} v_{i-1, 2},
\end{eqnarray}
where $T_{a, b}$ is the $(a, b)$-th component of the transfer matrix $T(\mathbf{c}_{i}, \mathbf{c}_{i+1})$. This mapping depends only on the first two rows of the $2\times W$ configuration. If the first two rows are (h, h)-pairs, the constraint maps to itself. If the column configuration has any other type, the constraint maps to a new constraint. For $W=3$ configurations, this new constraint is not one of the four constraints [Eqs.~\eqref{eq: w3 const 1}-\eqref{eq: w3 const 4}] we find in the main text. When this mapped constraint is taken together with one of the four constraints, the mapped constraint results in either the top two rows, or bottom two rows to deform as an area of CRS, breaking the NT condition. This is not a valid $W=3$ strip mode. Thus the constraint $v_{i, 1} = -v_{i, 2}$ is incompatible with a valid strip deformation for $W=3$ strip modes.

\section{Diagonal compatibility constraints}
Here we derive the diagonal compatibility constraint [Eq.~\eqref{eq: plaquette constraint}] for all $2\times 2$ configurations $(c_{i, j}, c_{i, j+1}, c_{i+1, j}, c_{i+1, j+1})$. We consider all configurations one-by-one. We first consider the configurations for which every diagonal edge $d$ in the diagonal compatibility constraint [Eq.~\eqref{eq: plaquette constraint}] is trivially zero. The diagonal compatibility constraint is then trivially satisfied and imposes no condition on $\mathbf{v}_i$.

Subsequently, we consider the case where $d_{i, j}^{\mathrm{SE}}$ is the only nontrivial diagonal edge in the diagonal compatibility constraint. To satisfy this constraint, we require $d_{i, j}^{\mathrm{SE}} = 0$ to hold. From Eq.~\eqref{eq: d in u, v} we find that this constraint only holds if $(i, j)$ is a CRS site, such that $u_{i, j} = v_{i, j}$. Similarly, if $d_{i+1, j}^{\mathrm{SW}}$ is the only nontrivial diagonal edge, we find $u_{i+1, j} = v_{i+1, j}$, if $d_{i, j+1}^{\mathrm{NE}}$ is the only nontrivial diagonal edge, we find $u_{i, j+1} = v_{i, j+1}$, and if $d_{i+1, j+1}^{\mathrm{NW}}$ is the only nontrivial diagonal edge, we find $u_{i+1, j+1} = v_{i+1, j+1}$. Thus a $2\times 2$ configuration where a single building block is oriented such that its nontrivial diagonal edge $d^c$ is part of the diagonal compatibility constraint [Eq.~\eqref{eq: plaquette constraint}] must deform that single building block as a CRS site to satisfy the diagonal compatibility constraint.

Now we consider configurations that contain horizontally paired building blocks which have nontrivial diagonal edges $d^c$ that are both part of the diagonal compatibility constraint [Eq.~\eqref{eq: plaquette constraint}]. The two other building blocks' diagonal edges in the diagonal compatibility constraint are trivial. We first consider configurations where such a pair of building blocks is in the top row, such that $(c_{i, j}, c_{i+1, j}) = (SE, SW)$. The diagonal compatibility constraint reduces to 
\begin{eqnarray}
    d_{i, j}^{\mathrm{SE}} + d_{i+1, j}^{\mathrm{SW}} = 0 \nonumber\\
    u_{i, j} - v_{i, j} + u_{i+1, j} - v_{i+1, j} = 0,
\end{eqnarray}
where we used Eq.~\eqref{eq: d in u, v} to replace $d^c$. We can simplify this further by replacing $v_{i+1, j}$ using the map Eq.~\eqref{eq: v_i+1 to v_i map}:
\begin{eqnarray}
    \left(1+\frac{R_u(c_{i, j})}{L_v(c_{i+1, j})}\right)v_{i, j-1} + \left(1-\frac{R_v(c_{i, j})}{L_v(c_{i+1, j} )}\right)v_{i, j} \nonumber\\
    + \left(1+\frac{L_u(c_{i+1, j})}{L_v(c_{i+1, j})}\right)v_{i+1, j-1} = 0,
\end{eqnarray}
where we have also used the horizontal compatibility constraint to replace $u$ with $v$. Replacing the components $R_u, L_u, L_v$ by their explicit values for $(c_{i, j}, c_{i+1, j}) = (SE, SW)$ (Table~\ref{tab:L, R, D}) we finally find the constraint
\begin{equation}
    v_{i, j-1} = -v_{i+1, j-1}.
    \label{eq: (SE, SW) hor pair diag constr}
\end{equation}
Similarly, we can derive the constraint for when the horizontally paired building are located in the bottom row, such that $(c_{i, j+1}, c_{i+1, j+1}) = (NE, NW)$. We find the constraint
\begin{equation}
    v_{i, j} = - v_{i+1, j}.
    \label{eq: (NE, NW) hor pair diag constr}
\end{equation}
Thus we find that such horizontally paired building blocks must deform their upper edges anti-symmetrically to satisfy the diagonal compatibility constraint. We can write Eq.~\eqref{eq: (NE, NW) hor pair diag constr} as a constraint on $v_{i, j-1}$ by replacing $v_{i+1, j}$ using map [Eq.~\eqref{eq: v_i+1 to v_i map}] and find
\begin{eqnarray}
    v_{i, j-1} + v_{i+1, j-1} = 0, \label{eq: (h, h) hor (NE, NW) diag constr}\\
    2 v_{i, j} +  v_{i, j-1} - v_{i+1, j-1} = 0, \label{eq: (u, h) NW hor (NE, NW) diag constr} \\
    \frac{2}{3} v_{i, j} - v_{i, j-1} + v_{i+1, j-1} = 0 \label{eq: (u, h) NE hor (NE, NW) diag constr}
\end{eqnarray}
for (h, h)-pairs and (u, h)-pairs with $c_{i+1, j} = NW$ or $c_{i, j} = NE$ respectively.

Now we consider configurations which contain vertically paired building blocks which have nontrivial diagonal edges that both are part of the same diagonal compatibility constraint [Eq.~\eqref{eq: plaquette constraint}]. The other two building blocks' diagonal edges in the diagonal compatibility constraint are trivial. We first consider configurations where this vertical pair of building blocks is located on the left, such that $(c_{i, j}, c_{i, j+1}) = (SE, NE)$. The diagonal compatibility constraint reduces to
\begin{eqnarray}
    d_{i, j}^{SE} + d_{i, j+1}^{NE} = 0 \nonumber\\
    u_{i, j} - v_{i, j} - u_{i, j+1} + v_{i, j+1} = 0 \nonumber\\
    -v_{i, j-1} + v_{i, j+1} = 0,
    \label{eq: (SE, NE) ver par diag constr}
\end{eqnarray}
where we used the horizontal compatibility constraint [Eq.~\eqref{eq: compatibility constraints}] to replace $u$ with $v$. 
Now we consider configurations where this vertical pair is located on the right, such that $(c_{i+1, j}, c_{i+1, j+1}) = (SW, NW)$. The diagonal compatibility constraint reduces to
\begin{equation}
    -v_{i+1, j-1} + v_{i+1, j+1} = 0
\end{equation}
We now try to map this constraint on $\mathbf{v}_{i+1}$ to a constraint on $\mathbf{v}_i$ using map [Eq.~\eqref{eq: v_i+1 to v_i map}]. We find that the constraint maps to
\begin{eqnarray}
    \left(\frac{L_u(c_{i+1, j+1}) L_u(c_{i+1, j})}{L_v(c_{i+1, j+1}) L_v(c_{i+1, j})} -1\right) v_{i+1, j-1} \nonumber\\
    + \left(\frac{R_u(c_{i, j+1})}{L_v(c_{i+1, j+1})} - \frac{L_u(c_{i+1, j+1}) R_v(c_{i, j})}{L_v(c_{i+1, j+1}) L_v(c_{i+1, j})}\right)v_{i, j} \nonumber\\
    - \frac{R_v(c_{i, j+1})}{L_v(c_{i+1, j+1})} v_{i, j+1} \nonumber\\
    + \frac{L_u(c_{i+1, j+1}) R_u(c_{i, j})}{L_v(c_{i+1, j+1}) L_v(c_{i+1, j})} v_{i, j-1} = 0
\end{eqnarray}
Depending on the precise orientations of the building blocks in the $2\times 2$ configuration, this constraint reduces to two different constraints (Table~\ref{tab:L, R, D})
\begin{eqnarray}
    v_{i, j+1} - v_{i, j-1} = 0 \label{eq: ver par diag constr no or two s pair}\\
    \frac{2}{3} v_{i, j} + \frac{1}{3} v_{i, j+1} + \frac{1}{3} v_{i, j-1} = 0. \label{eq: ver par diag constr single s pair}
\end{eqnarray}
The first constraint [Eq.~\eqref{eq: ver par diag constr no or two s pair}] must hold to satisfy the diagonal compatibility constraint for (d, u)-pairs and (s, \reverses)-pairs. The second constraint [Eq.~\eqref{eq: ver par diag constr single s pair}] must hold to satisfy the diagonal compatibility constraint for (d, \reverses)-pairs and (s, u)-pairs.

Next we consider configurations which contain diagonally paired building blocks which have nontrivial diagonal edges that are part of the same diagonal compatibility constraint. The other two building block's diagonal edges in the diagonal compatibility constraint are trivial. We first consider the case where $(c_{i, j}, c_{i+1, j+1}) = (SE, NW)$. The diagonal compatibility constraint [Eq.~\eqref{eq: plaquette constraint}] reduces to
\begin{eqnarray}
    d_{i, j}^{SE} + d_{i+1, j+1}^{NW} = 0 \\
    -v_{i, j-1} + v_{i, j} + v_{i+1, j} + v_{i+1, j+1} = 0,
\end{eqnarray}
where we used Eq.~\eqref{eq: d in u, v} and the horizontal compatibility constraint to simplify the constraint. We use the map [Eq.~\eqref{eq: v_i+1 to v_i map}] to replace $\mathbf{v}_{i+1}$ by $\mathbf{v}_{i}$. We find four different constraints depending on the orientations of the building blocks in the $2\times 2$ configuration:
\begin{eqnarray}
    v_{i, j-1} -\frac{1}{3} v_{i, j+1} + \frac{2}{3} v_{i+1, j-1} = 0, \label{eq: (d, u) diag (SE, NW) diag constr}\\
    \frac{1}{3}v_{i, j-1} + \frac{4}{9}v_{i, j} + \frac{1}{3} v_{i, j+1} + \frac{2}{9} v_{i+1, j-1} = 0, \label{eq: (rs, u) diag (SE, NW) diag constr}\\
    v_{i, j-1} + \frac{2}{3} v_{i, j} + \frac{1}{3} v_{i, j+1} + \frac{2}{3} v_{i+1, j-1} = 0, \label{eq: (d, rs) diag (SE, NW) diag constr} \\
    -\frac{1}{3} v_{i, j-1} + \frac{2}{9} v_{i, j} + \frac{1}{3} v_{i, j+1} - \frac{2}{9} v_{i+1, j-1} = 0 \label{eq: (rs, rs) diag (SE, NW) diag constr},
\end{eqnarray}
for (d, u)-pairs, (\reverses, u)-pairs, (d, \reverses)-pairs, and (\reverses, \reverses)-pairs respectively. 
Second, we consider the case where $(c_{i+1, j}, c_{i, j+1}) = (SW, NE)$. The diagonal compatibility constraint reduces to
\begin{eqnarray}
    \left(1+\frac{R_v(c_{i, j})}{L_v(c_{i+1, j})}\right) v_{i, j} + v_{i, j+1} - \frac{R_u(c_{i, j})}{L_v(c_{i+1, j})} v_{i, j-1} \nonumber\\
    - \left( 1 + \frac{L_u(c_{i+1, j})}{L_v(c_{i+1, j})} \right) v_{i+1, j-1} = 0.
\end{eqnarray}
Depending on the orientations of the $2\times 2$ configuration, we find two different constraints:
\begin{eqnarray}
    v_{i, j+1} + v_{i, j-1} + 2 v_{i+1, j-1} = 0 \label{eq: (d, u) (d, s) diag (SW, NE) diag constr} \\
    -2 v_{i, j} + v_{i, j+1} - v_{i, j-1} + 2 v_{i+1, j-1} = 0 \label{eq: (s, u) (s, s) diag (SW, NE) diag constr}
\end{eqnarray}
for (d, u)-pairs and (d, s)-pairs, and (s, u)-pairs and (s, s)-pairs respectively.

Next, we consider configurations where three building blocks have a nontrivial diagonal edge that is part of the diagonal compatibility constraint [Eq.~\eqref{eq: plaquette constraint}]. We first consider $(c_{i, j}, c_{i+1, j}, c_{i, j+1}) = (\mathrm{SE, SW, NE})$. The diagonal compatibility constraint reduces to
\begin{eqnarray}
    v_{i, j-1} - v_{i, j+1} + v_{i+1, j-1} + v_{i+1, j} = 0.    
\end{eqnarray}
We replace $v_{i+1, j}$ using map [Eq.~\eqref{eq: v_i+1 to v_i map}] for $(c_{i, j}, c_{i+1, j}) = (SE, SW)$ to find the constraint
\begin{eqnarray}
    -2 v_{i, j-1} - v_{i, j+1} - v_{i, j} -2 v_{i+1, j-1} = 0,
    \label{eq: (h, u) (h, s) triplet (SE, SW, NE) diag constr}
\end{eqnarray}
regardless of the orientation $c_{i+1, j+1}$. 
Next we consider $(c_{i, j}, c_{i+1, j}, c_{i+1, j+1}) = (SE, SW, NW)$. The diagonal compatibility constraint reduces to
\begin{eqnarray}
    v_{i, j-1} +v_{i, j} + v_{i+1, j-1} - v_{i+1, j+1} = 0.
\end{eqnarray}
We replace $v_{i+1, j+1}$ using the map [Eq.~\eqref{eq: v_i+1 to v_i map}] and find
\begin{eqnarray}
    v_{i, j-1} + \left(1 - \frac{R_u(c_{i, j+1})}{L_v(c_{i+1, j+1})}\right)v_{i, j} \nonumber\\ + \frac{R_v(c_{i, j+1})}{L_v(c_{i+1, j+1})} v_{i, j+1} 
     + v_{i+1, j-1} \nonumber\\ + \frac{L_u(c_{i+1, j+1})}{L_v(c_{i+1,j+1})} v_{i+1, j} = 0
\end{eqnarray}
We again apply map [Eq.~\eqref{eq: v_i+1 to v_i map}] to replace $v_{i+1, j}$ and find the same constraint regardless of the orientation $c_{i, j+1}$:
\begin{eqnarray}
    v_{i, j} + v_{i, j+1} = 0.
    \label{eq: (h, u) (h, rs) triplet (SE, SW, NW) diag constr}
\end{eqnarray}
This constraint requires the building block at site $(i, j+1)$ to deform as a CRS block to satisfy the diagonal compatibility constraint.
Next we consider $(c_{i, j}, c_{i, j+1}, c_{i+1, j+1}) = (\mathrm{SE, NE, NW})$. The diagonal compatibility constraint reduces to
\begin{equation}
    v_{i, j-1} - v_{i,k j+1} - v_{i+1, j} - v_{i+1, j+1} = 0.
\end{equation}
We replace $\mathbf{v}_{i+1}$ using the map [Eq.~\eqref{eq: v_i+1 to v_i map}] and find one of two constraints
\begin{eqnarray}
    -v_{i, j-1} + \frac{1}{3} v_{i, j} - \frac{2}{3} v_{i+1, j-1} = 0 \label{eq: (d, h) triplet (SE, NE, NW) diag constr} \\
    v_{i, j-1} + \frac{1}{3} v_{i, j} + \frac{2}{3} v_{i+1, j-1} = 0, \label{eq: (rs, h) triplet (SE, NE, NW) diag constr}
\end{eqnarray}
for a (d, h)-pair or (\reverses, h)-pair respectively. 
Finally, we consider the configuration $(c_{i+1, j}, c_{i, j+1}, c_{i+1, j+1}) = (\mathrm{SW, NE, NW})$. The diagonal compatibility constraint reduces to 
\begin{equation}
    -v_{i, j} - v_{i, j+1} + v_{i+1, j-1} - v_{i+1, j+1} = 0.
\end{equation}
We replace $\mathbf{v}_{i+1}$ using map [Eq.~\eqref{eq: v_i+1 to v_i map}] and find the constraint
\begin{equation}
    v_{i, j-1} + v_{i, j} = 0
    \label{eq: (d, h) (s, h) triplet (SW, NE, NW) diag constr}
\end{equation}
regardless of the orientation of $c_{i, j}$. This constraint corresponds to site $(i, j)$ deforming as a CRS site to satisfy the diagonal compatibility constraint.

Finally, we consider the last $2\times 2$ configuration: $(c_{i, j}, c_{i+1, j}, c_{i, j+1}, c_{i+1, j+1}) = (\mathrm{SE, SW, NE, NW})$. The diagonal compatibility constraint reduces to
\begin{equation}
    v_{i, j-1} - v_{i, j+1} + v_{i+1, j-1} - v_{i+1, j+1} = 0.
\end{equation}
We replace $\mathbf{v}_{i+1}$ using map [Eq.~\eqref{eq: v_i+1 to v_i map}] to find
\begin{equation}
    v_{i, j-1} + v_{i+1, j-1} - v_{i, j-1} - v_{i+1, j-1} = 0,
\end{equation}
which is trivially true. Thus this $2\times 2$ configuration does not impose any additional constraints on $\mathbf{v}_i$ to satisfy the diagonal compatibility constraint.

\subsection{Diagonal constraints for $W=2$ configurations \label{App: Diag W2}}
These equations simplify further for specific cases. We first consider $W=2$ valid configurations and then $W=3$ valid configurations. For $W=2$ strip configurations, $1\leq j \leq 2$ and $v_{i, 0} = -v_{i+1, 0} = 0$. Moreover, the lower boundary condition dictates $v_{i, 2} = -v_{i+1, 2}$. We further simplify the diagonal compatibility constraint [Eq.~\eqref{eq: plaquette constraint}] by assuming the lower and upper boundary conditions are satisfied.

We find for (h, h)-pairs that constraint [Eq.~\eqref{eq: (NE, NW) hor pair diag constr}] is trivially satisfied by the top boundary condition. Thus (h, h)-pairs do not impose any additional constraints on the strip deformation. Additionally, (d, u)-pairs can impose the constraints [Eqs.~\eqref{eq: ver par diag constr no or two s pair}, \eqref{eq: (d, u) diag (SE, NW) diag constr}, and \eqref{eq: (d, u) (d, s) diag (SW, NE) diag constr}], which simplify to
\begin{eqnarray}
    v_{i, 2} = 0
    \label{eq: w2 valid config constr}
\end{eqnarray}
for all cases.

Now, we consider (h, u)-pairs and (d, h)-pairs. The constraints [Eqs.~\eqref{eq: (h, u) (h, s) triplet (SE, SW, NE) diag constr},  \eqref{eq: (h, u) (h, rs) triplet (SE, SW, NW) diag constr},  \eqref{eq: (d, h) triplet (SE, NE, NW) diag constr}, and \eqref{eq: (d, h) (s, h) triplet (SW, NE, NW) diag constr}] reduce to
\begin{eqnarray}
    v_{i, 1} + v_{i, 2} = 0, \quad \mathrm{or} \\
    v_{i, 1} = 0
\end{eqnarray}
for (h, u)-pairs and (d, h)-pairs respectively.
Both of these constraints are not compatible with a strip deformation as they break the nontrivial (NT) condition. 

\subsection{Diagonal constraints for $W=3$ configurations \label{App: Diag W3}}
Here, we consider the diagonal constraints for $W=3$ configurations. Valid $3\times 2$ configurations are (h, h, h)-pairs, (d, u, h)-pairs, (h, d, u)-pairs, (h, s, u)-pairs and (h, \reverses, u)-pairs. The upper boundary condition implies $v_{i, 0} = -v_{i+1, 0} = 0$ and the bottom boundary condition implies $v_{i, 3} = -v_{i+1, 3} = 0$. We consider each configuration one-by-one.

We first consider (h, h, h)-pairs, which can only impose constraint [Eq.~\eqref{eq: (SE, SW) hor pair diag constr}] for all $j$. It is straightforward to check that the map [Eq.~\eqref{eq: v_i+1 to v_i map}] for (h, h, h)-pairs trivially satisfies all possible diagonal compatibility constraints. 

Second, we consider (d, u, h)-pairs, which can impose the constraints [Eqs.~\eqref{eq: (u, h) NE hor (NE, NW) diag constr}, \eqref{eq: (u, h) NW hor (NE, NW) diag constr}, \eqref{eq: (d, u) (d, s) diag (SW, NE) diag constr}, and \eqref{eq: (d, u) diag (SE, NW) diag constr}].  Combining the possible constraints together with the upper and bottom boundary conditions and map [Eq.~\eqref{eq: v_i+1 to v_i map}] impose the constraint
\begin{eqnarray}
    v_{i, 2} = 0.
\end{eqnarray}
Thus (d, u, h)-pairs impose constraint [Eq.~\eqref{eq: w3 const 1}] on $\mathbf{v}_i$ to satisfy the diagonal compatibility constraints.

Third, we consider (h, d, u)-pairs. Such pairs can impose the constraints [Eqs.~\eqref{eq: (SE, SW) hor pair diag constr}, \eqref{eq: (d, u) (d, s) diag (SW, NE) diag constr}, and \eqref{eq: (d, u) diag (SE, NW) diag constr}], which simplify to
\begin{equation}
    v_{i, 1} = v_{i, 3} 
\end{equation}
using the upper and lower boundary conditions. Thus (h, d, u)-pairs impose constraint [Eq.~\eqref{eq: w3 const 3}] to satisfy the diagonal compatibility constraints.

Fourth, we consider (d, s, u)-pairs. Such pairs impose the constraints [Eqs.~\eqref{eq: ver par diag constr single s pair}, \eqref{eq: (d, u) (d, s) diag (SW, NE) diag constr}, and \eqref{eq: (s, u) (s, s) diag (SW, NE) diag constr}], which simplify to the constraints
\begin{eqnarray}
    v_{i, 2} = 0, \quad \mathrm{and} \\
    v_{i, 1} = - v_{i, 3}.
\end{eqnarray}
Thus (d, s, u)-pairs impose constraints [Eqs.~\eqref{eq: w3 const 1} and \eqref{eq: w3 const 4}] to satisfy the diagonal compatibility constraints.

Finally, we consider (d, \reverses, u)-pairs. Such pairs impose the constraints [Eqs.~\eqref{eq: ver par diag constr single s pair}, \eqref{eq: (d, rs) diag (SE, NW) diag constr}, and \eqref{eq: (rs, u) diag (SE, NW) diag constr}], which simplify to the constraints
\begin{eqnarray}
    2 v_{i, 1} = -v_{i, 2}, \quad \mathrm{and} \\
    v_{i, 1} = v_{i, 3}.
\end{eqnarray}
Thus (d, \reverses, u)-pairs impose constraints [Eqs.~\eqref{eq: w3 const 2} and \eqref{eq: w3 const 3}] to satisfy the diagonal compatibility constraints.

In summary, we find that (h, h, h)-pairs do not impose any constraints on the strip deformation $\mathbf{v}_i$, (d, u, h)-pairs impose constraint [Eq.~\eqref{eq: w3 const 1}], (h, d, u)-pairs impose constraint [Eq.~\eqref{eq: w3 const 3}], (d, s, u)-pairs impose constraints [Eqs.~\eqref{eq: w3 const 1} and \eqref{eq: w3 const 4}], and (s, \reverses, u)-pairs impose constraints [Eqs.~\eqref{eq: w3 const 2} and \eqref{eq: w3 const 3}].

Now we consider invalid configurations that contain diagonal compatibility constraints with three nontrivial diagonal edges. First, we consider (h, u, h)-pairs. Such pairs impose the constraints [Eqs.~\eqref{eq: (h, u) (h, rs) triplet (SE, SW, NW) diag constr} and \eqref{eq: (h, u) (h, s) triplet (SE, SW, NE) diag constr}], which simplify to
\begin{equation}
    v_{i, 1} = -v_{i, 2}.
\end{equation}
This corresponds to the block at site $(i, 1)$ deforming as a CRS block, which is incompatible with a $W=3$ strip deformation.
Second, we consider (d, h, h)-pairs. Such pairs impose the constraints [Eqs.~\eqref{eq: (d, h) (s, h) triplet (SW, NE, NW) diag constr} and \eqref{eq: (d, h) triplet (SE, NE, NW) diag constr}], which simplify to
\begin{equation}
    v_{i, 1} = 0,
\end{equation}
which is incompatible with a valid strip deformation.
Third we consider (h, s, h)-pairs. Such pairs impose the constraints [Eqs.~\eqref{eq: (h, u) (h, s) triplet (SE, SW, NE) diag constr} and \eqref{eq: (d, h) (s, h) triplet (SW, NE, NW) diag constr}], which simplify to
\begin{equation}
    v_{i, 2} = -v_{i, 1},
\end{equation}
which is incompatible with a valid strip deformation.
Fourth, we consider (h, \reverses, h)-pairs. Such pairs impose the constraints [Eqs.~\eqref{eq: (h, u) (h, rs) triplet (SE, SW, NW) diag constr} and \eqref{eq: (rs, h) triplet (SE, NE, NW) diag constr}], which simplify to
\begin{equation}
    v_{i, 1} = -v_{i, 2},
\end{equation}
which is incompatible with a valid strip deformation. (h, s, u)-pairs, (h, \reverses, u)-pairs, (d, s, h)-pairs and (d, \reverses, h)-pairs all impose the same constraint on the strip deformation.
Finally, we consider (h, h, u)-pairs. Such pairs impose the constraints [Eqs.~\eqref{eq: (h, u) (h, rs) triplet (SE, SW, NW) diag constr} and \eqref{eq: (h, u) (h, s) triplet (SE, SW, NE) diag constr}], which simplify to
\begin{eqnarray}
    v_{i, 2} = -v_{i, 3}.
\end{eqnarray}
This is incompatible with a valid strip deformation.

\section{lower strip condition \label{App: lower strip condition}}
In this appendix we find constraints that need to be satisfied in order to satisfy the lower boundary condition [Eq.~\eqref{eq: BC bottom}]. We first consider $W=2$ configurations and then $W=3$ configurations. We find that for valid configurations the lower boundary condition is satisfied by the same constraints that arise from the diagonal compatibility constraints [Eq.~\eqref{eq: plaquette constraint}]. 

First, we consider $W=2$ configurations. In general, the map [Eq.~\eqref{eq: W2 map}] does not satisfy the lower boundary condition $v_{i+1, 2} = -v_{i, 2}$. We first consider (h, h)-pairs. Here the map [Eq.~\eqref{eq: W2 map}] reduces to $v_{i+1, 2} = -v_{i, 2}$, and the lower boundary condition is trivially satisfied. Next, we consider (d, u)-pairs. The map [Eq.~\eqref{eq: W2 map}] together with the lower boundary condition reduces to the constraint
\begin{eqnarray}
    \left( 1-\frac{R_v(c_{i, 2})}{L_v(c_{i+1, 2})} \right) v_{i, 2} = 0.
\end{eqnarray}
For (d, u)-pairs, $R_v(c_{i, 2}) / L_v(c_{i+1, 2})$ is either equal to $3$ or $1 / 3$. This constraint is thus satisfied if $v_{i, 2} = 0$, which is the same as the constraint [Eq.~\eqref{eq: w2 valid config constr}] needed to satisfy the diagonal compatibility constraint. Thus the lower boundary condition and diagonal compatibility constraint for (d, u)-pairs both are satisfied if constraint $v_{i, 2} = 0$ is satisfied.

Now, we consider $W=3$ configurations. First, we consider (h, h, h)-pairs. The map [Eq.~\eqref{eq: W3 map}] reduces to $v_{i+1, 3} = -v_{i, 3}$, thus trivially satisfying the lower boundary condition [Eq.~\eqref{eq: plaquette constraint}]. Second, we consider (d, u, h)-pairs. The map [Eq.~\eqref{eq: W3 map}] together with the lower boundary condition [Eq.~\eqref{eq: plaquette constraint}] reduces to the constraint
\begin{equation}
    \frac{L_u (c_{i+1, 3})}{L_v(c_{i+1, 3})} \left( 1 - \frac{R_v(c_{i, 2})}{L_v(c_{i+1, 2})} \right) v_{i, 2} = 0,
\end{equation}
which is satisfied by the constraint [Eq.~\eqref{eq: w3 const 1}], just like the diagonal compatibility constraint for (d, u, h)-pairs.
Third, we consider (h, d, u)-pairs. The map [Eq.~\eqref{eq: W3 map}] together with the lower boundary condition reduces to the constraint
\begin{eqnarray}
    \frac{L_u(c_{i+1, 2}) L_u(c_{i+1, 3})}{L_v(c_{i+1, 2}) L_v(c_{i+1, 3})} \left( \frac{R_u(c_{i, 2})}{L_v(c_{i+1, 2})} - 1 \right) v_{i, 1} \nonumber\\
    + \left(1 - \frac{R_v(c_{i, 3})}{L_v(c_{i+1, 3})} v_{i, 3} \right) v_{i, 3} = 0.
\end{eqnarray}
This constraint reduces to $v_{i, 1} = v_{i, 3}$ by filling in the possible explicit values of $(c_{i, 2}, c_{i+1, 2}, c_{i, 3}, c_{i+1, 3})$. Thus the lower boundary condition and diagonal compatibility constraint are satisfied by satisfying constraint [Eq.~\eqref{eq: w3 const 3}] for (h, d, u)-pairs.

Now, we consider (d, s, u)-pairs. The map [Eq.~\eqref{eq: W3 map}] together with the lower boundary condition [Eq.~\eqref{eq: BC bottom}] reduces to
\begin{equation}
    -v_{i, 1} + v_{i, 2} - v_{i, 3} = 0.
\end{equation}
This equation is satisfied by constraints [Eqs.~\eqref{eq: w3 const 1} and \eqref{eq: w3 const 4}], just like the diagonal compatibility constraints [Eq.~\eqref{eq: plaquette constraint}] for (d, s, u)-pairs.

Finally, we consider (d, \reverses, u)-pairs. The map [Eq.~\eqref{eq: W3 map}] together with the lower boundary condition [Eq.~\eqref{eq: BC bottom}] reduces to the constraint 
\begin{equation}
    v_{i, 1} - v_{i, 2} - 3 v_{i, 3} = 0.
\end{equation}
This constraint is satisfied by constraints [Eqs.~\eqref{eq: w3 const 2} and \eqref{eq: w3 const 3}]. Thus the lower boundary condition and diagonal compatibility constraints for (d, \reverses, u)-pairs are satisfied by constraints [Eqs.~\eqref{eq: w3 const 2} and \eqref{eq: w3 const 3}].
In conclusion, we find that satisfying the lower boundary constraints does not require any additional constraints than required to satisfy the diagonal compatibility constraints for valid $W=3$ configurations: (h, h, h)-pairs, (d, u, h)-pairs, (h, d, u)-pairs, (d, s, u)-pairs and (d, \reverses, u)-pairs.

\section{Constraint mapping \label{App: constraint mapping}}
In general, a linear constraint $f(\mathbf{v}_i, \mathbf{v}_{i+1}) = g(\mathbf{v}_i) + h(\mathbf{v}_{i+1}) =  0$ depends on the deformations of two adjacent columns: $\mathbf{v}_i$ and $\mathbf{v}_{i+1}$. To map this constraint to a constraint on $\mathbf{v}_1$, we iteratively apply the transfer matrix to obtain a constraint on $\mathbf{v}_1$
\begin{eqnarray}
    \prod_{a=1}^{i-1}T(\mathbf{c}_{i-a}, \mathbf{c}_{i+1-a}) g(\mathbf{v}_1) \nonumber\\
    + \prod_{a=1}^{i}T(\mathbf{c}_{i+1-a}, \mathbf{c}_{i+2-a}) h(\mathbf{v}_1) = 0
\end{eqnarray}

\section{$W=3$ constraint mapping \label{App: constraint mapping W3}}
To find which combinations of valid $2\times 3$ configurations lead to $W=3$ strip modes, we consider how the four constraints [Eqs.~\eqref{eq: w3 const 1}-\eqref{eq: w3 const 4}] on $\mathbf{v}_{i}$ map to $\mathbf{v}_{i+1}$ upon application of the transfer matrix $T(\mathbf{c}_{i-1}, \mathbf{c}_i)$ for all possible valid $2\times 3$ configurations $(\mathbf{c}_{i-1}, \mathbf{c}_i)$.  We first derive the transfer matrices for each valid configuration, which we subsequently apply tot the four constraints to see how they map.

First, we consider the transfer matrix for (h, h, h)-pairs: $T((h, h, h))$. We find that $T((h, h, h)) = - I$, where $I$ is the $3\times 3$ identity matrix. Thus the four constraints on $\mathbf{v}_{i}$ map to the same constraints on $\mathbf{v}_{i-1}$.

Second, we consider the transfer matrix for (d, u, h)-pairs. We find
\begin{equation}
    T((d, u, h)) = \begin{pmatrix}
        1   &   0   &   0   \\
        0   &   -\frac{R_v(c_{i-1, 2})}{L_v(c_{i, 2})}  &   0   \\
        0   &  \frac{L_u(c_{i, 3})}{L_v(c_{i, 3})} \left( 1 - \frac{R_v(c_{i-1, 2})}{L_v(c_{i, 2})} \right) &  -1  \\
    \end{pmatrix},
    \label{eq: T matrix (d, u, h)}
\end{equation}
where the precise orientations of $(c_{i-1, 2}, c_{i, 2})$ determine the explicit form of the matrix. Now we apply this transfer matrix on the four constraints and rewrite the constraints together with constraint [Eq.~\eqref{eq: w3 const 1}] on $\mathbf{v}_{i-1}$ imposed by the (d, u, h)-pair itself. We find that constraint [Eq.~\eqref{eq: w3 const 1}] maps to itself with $i\mapsto i-1$. Constraint [Eq.~\eqref{eq: w3 const 2}] maps to $v_{i-1, 1} = 0$, which is incompatible with a $W=3$ strip deformation as it breaks the nontrivial (NT) condition. Finally, constraint [Eq.~\eqref{eq: w3 const 3}] maps to constraint [Eq.~\eqref{eq: w3 const 4}] with $i\mapsto i-1$ and similarly constraint [Eq.~\eqref{eq: w3 const 3}] maps to constraint [Eq.~\eqref{eq: w3 const 4}] with $i\mapsto i-1$.

Third, we consider the transfer matrix for (h, d, u)-pairs:
\begin{equation}
    T((h, d, u)) = \begin{pmatrix}
        -1      &       0       &       0       \\
        2       &       1       &       0       \\
         \frac{L_u(c_{i, 3})}{L_v(c_{i, 3})} 2      &       0       &       -\frac{R_v(c_{i-1, 3})}{L_v(c_{i, 3})}
    \end{pmatrix},
    \label{eq: T matrix (h, d, u)}
\end{equation}
which depends on the precise orientations of $(c_{i-1, 3}, c_{i, 3})$. If we apply this transfer matrix on the four constraints and take the mapped constraint together with the constraint [Eq.~\eqref{eq: w3 const 3}[] imposed by the (h, d, u)-pair on $\mathbf{v}_{i-1}$ we find that constraint [Eq.~\eqref{eq: w3 const 1}] maps to constraint [Eq.~\eqref{eq: w3 const 2}] with $i\mapsto i-1$. Constraint [Eq.~\eqref{eq: w3 const 2}] maps to constraint [Eq.~\eqref{eq: w3 const 1}] with $i\mapsto i-1$. Constraint [Eq.~\eqref{eq: w3 const 3}] maps to itself with $i\mapsto i-1$. Constraint [Eq.~\eqref{eq: w3 const 4}] maps to $v_{i-1, 1} = v_{i-1, 3} = 0$ when taken together with constraint [Eq.~\eqref{eq: w3 const 3}] with $i\mapsto i-1$ imposed by the (h, d, u)-pair, which is incompatible with a $W=3$ strip deformation as it breaks the NT condition.

Fourth, we consider the transfer matrix for (d, s, u)-pairs:
\begin{eqnarray}
    T((d, s, u)) = \nonumber\\ 
    \begin{pmatrix}
        1               &               0               &           0               \\
       - 2      &       3          &       0           \\
        - 2 \frac{L_u(c_{i, 3})}{L_v(c_{i, 3})}            &     2   \frac{L_u(c_{i, 3})}{L_v(c_{i, 3})}    &       -\frac{R_v(c_{i-1, 3})}{L_v(c_{i, 3})} 
    \end{pmatrix} ,
    \label{eq: T matrix (d, s, u)}
\end{eqnarray}
which depends on the orientations of $(c_{i-1, 3}, c_{i, 3})$. We find that constraint [Eq.~\eqref{eq: w3 const 1}] maps to the constraint $v_{i-1, 1} = 0$ when taken together with constraint [Eq.~\eqref{eq: w3 const 1}] with $i\mapsto i-1$ imposed by the (d, s, u)-pair itself. This constraint $v_{i-1, 1} = 0$ is incompatible with a valid strip deformation. Constraint [Eq.~\eqref{eq: w3 const 2}] maps to constraint [Eq.~\eqref{eq: w3 const 2}] with $i\mapsto i-1$. Constraint [Eq.~\eqref{eq: w3 const 3}] maps to constraint [Eq.~\eqref{eq: w3 const 4}] with $i\mapsto i-1$ when considered together with constraint [Eq.~\eqref{eq: w3 const 1}] with $i\mapsto i-1$ imposed by the (d, s, u)-pair. Finally, constraint [Eq.~\eqref{eq: w3 const 4}] maps to $v_{i-1, 1} = v_{i-1, 3} = 0$ when taken together with constraints [Eqs.~\eqref{eq: w3 const 1} and \eqref{eq: w3 const 4}] with $i\mapsto i-1$ imposed by the (d, s, u)-pair. This mapped constraint is not compatible with a valid strip deformation.

Finally, we consider the transfer matrix for (d, \reverses, u)-pairs:
\begin{eqnarray}
T((d, \reverses, u)) = \nonumber\\ 
    \begin{pmatrix}
        1               &               0               &           0               \\
       \frac{2}{3}     &      \frac{1}{3}         &       0           \\
        \frac{2}{3} \frac{L_u(c_{i, 3})}{L_v(c_{i, 3})}            &     -\frac{2}{3}  \frac{L_u(c_{i+1, 3})}{L_v(c_{i+1, 3})}    &       -\frac{R_v(c_{i-1, 3})}{L_v(c_{i, 3})} 
    \end{pmatrix} ,
    \label{eq: T matrix (d, rs, u)}
\end{eqnarray}
which depends on the orientations of $(c_{i-1, 3}, c_{i, 3})$. We find that constraint [Eq.~\eqref{eq: w3 const 1}] maps to constraint [Eq.~\eqref{eq: w3 const 2}] with $i\mapsto i-1$. Constraint [Eq.~\eqref{eq: w3 const 2}] maps to $v_{i-1, 1} = 0$ when taken together with constraint [Eq.~\eqref{eq: w3 const 2}] with $i\mapsto i-1$ imposed by the (d, \reverses, u)-pair itself. This mapped constraint is incompatible with a valid strip deformation. Constraint [Eq.~\eqref{eq: w3 const 3}] maps to $v_{i-1, 1} = v_{i-1, 3} = 0$ when taken together with constraints [Eqs.~\eqref{eq: w3 const 2} and \eqref{eq: w3 const 3}] with $i\mapsto i-1$ imposed by the (d, \reverses, u)-pair. This mapped constraint is incompatible with a valid strip deformation. Finally, constraint [Eq.~\eqref{eq: w3 const 4}] maps to constraint [Eq.~\eqref{eq: w3 const 3}] with $i\mapsto i-1$ using constraint [Eq.~\eqref{eq: w3 const 2}] with $i\mapsto i-1$.

\section{Invalid $W=3$ sequences \label{App: invalid sequences W3}}
We now explicitly state the combinations of valid $(\mathbf{c}_i, \mathbf{c}_{i+1})$ configurations that do not result in a valid $W=3$ strip mode.
The constraint mapping [Fig.~\ref{fig:w3 constraint mappings}(c)] and constraints imposed by the valid configurations prohibit certain combinations of valid configurations. In particular, we find for the top two rows in the strip that (d, u)-pairs or (d, s)-pairs cannot be preceded by a (d, s)-pair. Similarly, (d, \reverses)-pairs cannot be preceded by a (d, \reverses)-pair or (d, u)-pair. Additionally, a (d, u)-pair or (d, s)-pair preceded by a (h, d)-pair cannot be preceded by a (d, \reverses)-pair. Finally, a (d, \reverses)-pair preceded by a (h, d)-pair cannot be preceded by a (d, s)-pair. In the bottom two rows similar rules apply. There, we find that a (d, u)-pair or (\reverses, u)-pair cannot be preceded by a (\reverses, u)-pair. Similarly, a (s, u)-pair cannot be preceded by a (d, u)-pair or (s, u)-pair. Additionally, a (d, u)-pair or (d, \reverses)-pair preceded by a (u, h)-pair cannot be preceded by a (d, u)-pair or (s, u)-pair. Finally, a (s, u)-pair preceded by a (u, h)-pair cannot be preceded by a (\reverses, u)-pair. Sequences of such pairs in either the top or bottom two rows of the strip will not result in a valid strip mode. Note that such invalid sequences can be freely padded with (h, h)-pairs, as these map all the constraints to themselves and do not place any additional constraints on the strip deformation themselves. 

We now translate the invalid sequences of pairs to conditions on sequences of linked building blocks [Fig.~\ref{fig:linked rules}(a)]. First, we consider all invalid sequences of pairs in the top two rows. We note that all realizations of invalid sequences must feature a vertical and diagonal linked pair in the top two rows. To see this, note that (d, u)-pairs have one building block unlinked in the top row and one building block unlinked in the middle row, where these blocks are either both left or right if the two linked building blocks are linked vertically or the blocks are left and right or right and left if the two linked building blocks are linked diagonally. Similarly, (d, s)-pairs or (d, \reverses)-pairs have one building block unlinked within the two top rows: either to the left or right in the top row depending on if the two linked building blocks are diagonally or vertically linked ((d, s)-pairs), or vertically or diagonally linked ((d, \reverses)-pairs) respectively. Additionally, (h, d)-pairs have one unlinked building block in the top two rows have one unlinked building block depending on the orientation of the d-pair. To satisfy the strip rules, every building block needs to be linked.
To construct a linked representation that features vertical and diagonal linked building blocks necessarily requires an invalid sequence of pairs. Inversely, every realization of an invalid sequence of pairs necessarily requires vertical and diagonal linked building blocks.
The same holds for the invalid sequences in the bottom two rows. Thus we capture the exclusion of invalid sequences of pairs in a simple rule on linked building blocks in rule~\eqref{rule: w3 3}.

\section{Numerical proof strip mode rules \label{App: numerical proof rules}}
Here we show numerical proof that to have class (iii) mode-scaling, a unit cell
should support at least one strip mode. To show this, we use publicly available~\cite{Zenodo_MetaCombi} randomly generated $k\times k$ unit cell designs ranging from size $k=2$ to $k=8$ generated in earlier work~\cite{mastrigt2022machine}. We check for each generated unit cell if it obeys the strip mode rules as formulated in Sec.~\ref{sec:general design rules} using simple matrix operations and checks~\footnote{See \url{https://uva-hva.gitlab.host/published-projects/CombiMetaMaterial} for code to check the strip mode rules.}. We denote the number of strips in a unit cell that satisfy the strip mode rules as $a_{\mathrm{rule}}$ and compare this to the slope $a$ we find from the scaling of $N_{\mathrm{ZM}} =  a n + b$ . We do this for all possible $k=3$ unit cells and approximately one million $k=4, 5, 6$ unit cells, two million $k=7$ unit cells and 1.52 million $k=8$ unit cells.

\begin{figure*}[b]
    \centering
    \includegraphics{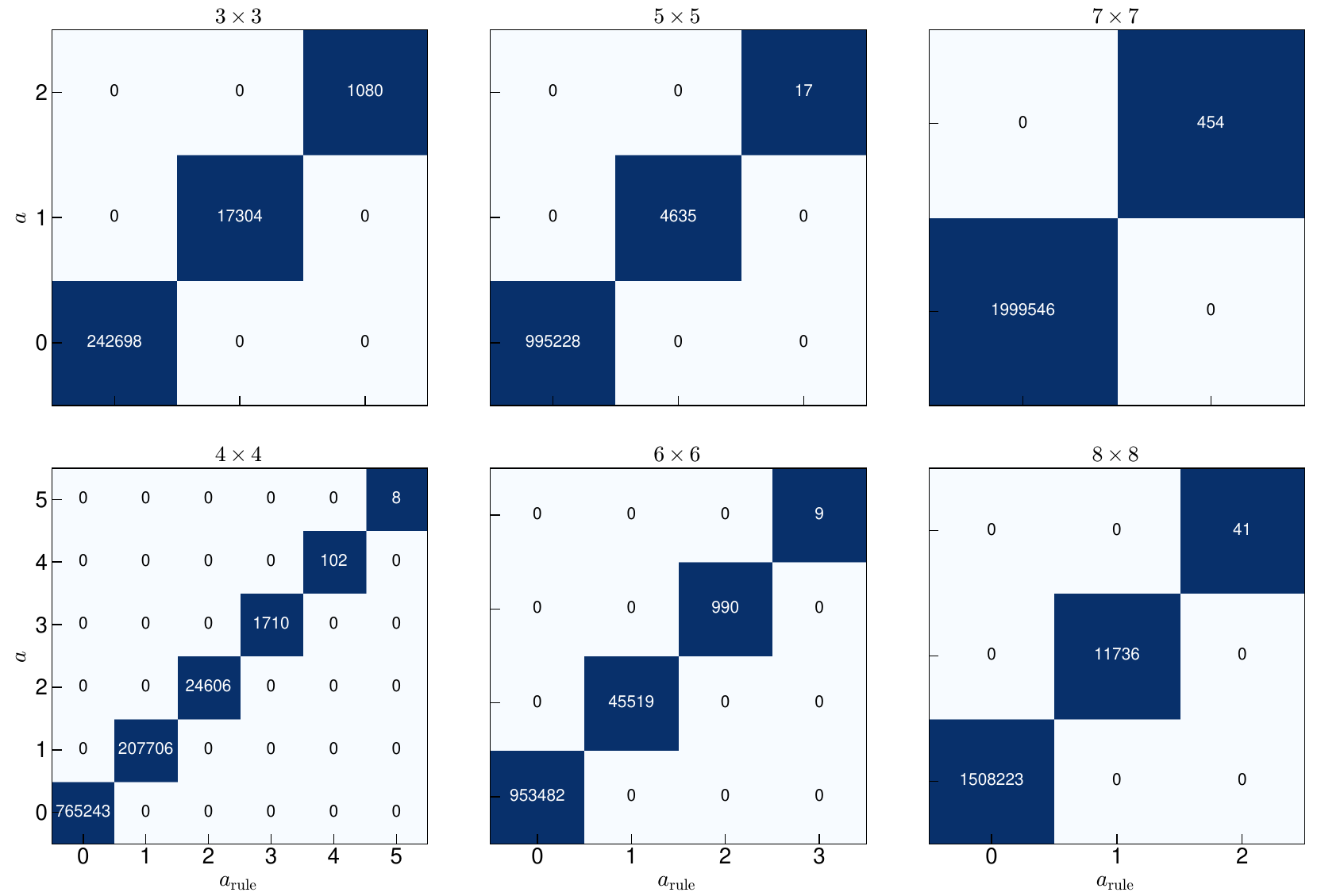}
    \caption{Confusion matrices comparing the slope $a$ based on mode-scaling $N_{\mathrm{ZM}} (n)$ to the number of strips that obey the strip mode rules $a_{\mathrm{rule}}$ as formulated in Sec.~\ref{sec:general design rules}. The $k \times k$ unit cell size is indicated on top of each matrix.}
    \label{fig:CM_rules}
\end{figure*}

We find that the true slope $a$ and the number of strip modes supported by the unit cell according to the strip mode rules $a_{\mathrm{rule}}$ have perfect agreement [Fig.~\ref{fig:CM_rules}].
The numerical results thus strongly suggest that supporting a strip mode is a necessary requirement for a unit cell to have class (iii) mode-scaling and that the conjectured general strip mode rules dictate when a unit cell supports such a strip mode. 

\end{document}